\documentclass[aps,prd,twocolumn,superscriptaddress]{revtex4}
\usepackage{epsfig,epsf}
\usepackage{amsmath}
\usepackage{amsthm}
\usepackage{amsfonts}
\usepackage{amssymb}
\usepackage{dsfont}
\usepackage{multirow}
\usepackage{appendix}
\usepackage{slashed}
\usepackage[active]{srcltx}
\usepackage{psfrag}

\begin{document}

\title{New charged resonance $Z_{c}^{-}(4100)$: the spectroscopic parameters
and width}
\date{\today}
\author{H.~Sundu}
\affiliation{Department of Physics, Kocaeli University, 41380 Izmit, Turkey}
\author{S.~S.~Agaev}
\affiliation{Institute for Physical Problems, Baku State University, Az--1148 Baku,
Azerbaijan}
\author{K.~Azizi}
\affiliation{Department of Physics, Do\v{g}u\c{s} University, Acibadem-Kadik\"{o}y, 34722
Istanbul, Turkey}
\affiliation{School of Particles and Accelerators, Institute for Research in Fundamental
Sciences (IPM) P.O. Box 19395-5531, Tehran, Iran}

\begin{abstract}
The mass, coupling and width of the newly observed charged resonance $%
Z_{c}^{-}(4100)$ are calculated by treating it as a scalar four-quark system
with a diquark-antidiquark structure. The mass and coupling of the state $%
Z_{c}^{-}(4100)$ are calculated using the QCD two-point sum rules. In these
calculations we take into account contributions of the quark, gluon and
mixed condensates up to dimension ten. The spectroscopic parameters of $%
Z_{c}^{-}(4100)$ obtained by this way are employed to study its $S$-wave
decays to $\eta_c(1S)\pi^{-}$, $\eta_c(2S)\pi^{-}$, $D^{0}D^{-}$, and $%
J/\psi \rho ^{-}$ final states. To this end, we evaluate the strong coupling
constants corresponding to the vertices $Z_{c}\eta_c(1S)\pi^{-}$, $%
Z_{c}\eta_c(2S)\pi^{-}$, $Z_{c}D^{0}D^{-}$, and $Z_{c}J/\psi \rho^{-}$
respectively. The couplings $g_{Z_c\eta_{c1} \pi}$, $g_{Z_{c}\eta_{c2} \pi}$%
, and $g_{Z_{c}DD}$ are computed by means of the QCD three-point sum rule
method, whereas $g_{Z_{c}J/\psi \rho }$ is obtained from the QCD light-cone
sum rule approach and soft-meson approximation. Our results for the mass $%
m=(4080 \pm 150)~\mathrm{MeV}$ and total width $\Gamma =(147 \pm 19)~\mathrm{%
MeV}$ of the resonance $Z_{c}^{-}(4100)$ are in excellent agreement with the
existing LHCb data.
\end{abstract}

\maketitle


\section{Introduction}

\label{sec:Intro}
Recently, the LHCb Collaboration reported on evidence for an $\eta
_{c}(1S)\pi ^{-}$ resonance in $B^{0}\rightarrow K^{+}\eta _{c}(1S)\pi ^{-}$
decays extracted from analysis of $pp$ collisions' data collected with LHCb
detector at center-of-mass energies of $\sqrt{s}=7,\ 8$ and $13\ \mathrm{TeV}
$ \cite{Aaij:2018bla}. The mass and width of this new $Z_{c}^{-}(4100)$
resonance (hereafter $Z_{c}$) were found equal to $m=4096\pm 20_{-22}^{ +18}~%
\mathrm{MeV}$ and $\Gamma =152\pm 58_{-35}^{+60}~\mathrm{MeV}$,
respectively. As it was emphasized in Ref. \cite{Aaij:2018bla}, the
spin-parity assignments $J^{P}=0^{+}$ and $J^{P}=1^{-}$ both are consistent
with the data.

From analysis of the decay channel $Z_{c}\rightarrow \eta _{c}(1S)\pi ^{-}$
it becomes evident that $Z_{c}$ contains four quarks $cd\overline{c}%
\overline{u}$, and it is presumably another member of the family of charged
exotic $Z$-resonances with the same quark content; the well-known
axial-vector tetraquarks $Z_{c}^{\pm }(4430)$ and $Z_{c}^{\pm }(3900)$ are
also built of the quarks $cd\overline{c}\overline{u}$ or $cu\overline{c}%
\overline{d}$. The $Z_{c}^{\pm }(4430)$ were discovered and studied by the
Belle Collaboration in $B$ meson decays $B\rightarrow K\psi ^{\prime }\pi
^{\pm }$ as resonances in the $\psi ^{\prime }\pi ^{\pm }$ invariant mass
distributions \cite{Choi:2007wga,Mizuk:2009da,Chilikin:2013tch}. The decay
of $Z_{c}^{+}(4430)$ $\ $to the final state $J/\psi \pi ^{+}$ also was
detected in the Belle experiment \cite{Chilikin:2014bkk}. The existence of
the $Z_{c}^{\pm }(4430)$ resonances was confirmed by the LHCb Collaboration
as well \cite{Aaij:2014jqa,Aaij:2015zxa}.

Another well-known members of this family are the axial-vector states $%
Z_{c}^{\pm }(3900)$, which were detected by the BESIII Collaboration in the
process $e^{+}e^{-}\rightarrow J/\psi \pi ^{+}\pi ^{-}$ as peaks in the $%
J/\psi \pi ^{\pm }$ invariant mass distributions \cite{Ablikim:2013mio}.
These structures were seen by the Belle and CLEO collaborations as well (see
Refs.\ \cite{Liu:2013dau,Xiao:2013iha}). The BESIII informed also on
observation of the neutral $Z_{c}^{0}(3900)$ state in the process $%
e^{+}e^{-}\rightarrow \pi ^{0}Z_{c}^{0}\rightarrow \pi ^{0}\pi ^{0}J/\psi $
\cite{Ablikim:2015tbp}.

Various theoretical models and approaches were employed to reveal the
internal quark-gluon structure and determine parameters of the charged $Z$%
-resonances. Thus, they were considered as hadrocharmonium compounds or
tightly bound diquark-antidiquark states, were treated as the four-quark
systems built of conventional mesons or interpreted as threshold cusps (see
Refs.\ \cite{Chen:2016qju,Esposito:2016noz} and references therein).

The diquark model of the exotic four-quark mesons is one of the popular
approaches to explain their properties. In accordance with this picture the
tetraquark is a bound state of a diquark and an antidiquark. This approach
implies the existence of multiplets of the diquark-antidiquarks with the
same quark content, but different spin-parities. Because the resonances $%
Z_{c}^{\pm }(3900)$ and $Z_{c}^{\pm }(4430)$ are the axial-vectors, one can
interpret them as the ground-state $1S$ and first radially excited $2S$
state of the same $[cu][\overline{c}\overline{d}]$ or $[cd][\overline{c}%
\overline{u}]$ multiplets. An idea to consider $Z_{c}(4430)$ as a radial
excitation of the $Z_{c}(3900)$ state was proposed in Ref.\ \cite%
{Maiani:2014}, and explored in Refs.\ \cite{Wang:2014vha,Agaev:2017tzv} in
the framework of the QCD sum rule method.

The resonances $Z_{c}(3900)$, $Z_{c}(4200)$, $Z_{c}(4430)$ and $Z_{c}$ were
detected in $B$ meson decays and/or electron-positron annihilations, which
suggest that all of them may have the same nature. Therefore, one can
consider $Z_{c}$ as the ground-state scalar or vector tetraquark with $c%
\overline{c}d\overline{u}$ content. The recent theoretical articles devoted
to the $Z_{c}$ resonance are concentrated mainly on exploration of its spin
and possible decay channels \cite%
{Wang:2018ntv,Wu:2018xdi,Voloshin:2018vym,Zhao:2018xrd,Cao:2018}. Thus, sum
rule computations carried out in Ref.\ \cite{Wang:2018ntv} demonstrated that
$Z_{c}$ is presumably a scalar particle rather than a vector tetraquark. The
conclusion about a tetraquark nature of $Z_{c}$ with quantum numbers $%
J^{PC}=0^{++}$ was drawn in Ref. \cite{Wu:2018xdi} as well. In the
hadrocharmonium framework the resonances $Z_{c}$ and $Z_{c}^{-}(4200)$ were
treated as the scalar $\eta _{c}$ and vector $J/\psi $ charmonia embedded in
a light-quark excitation with quantum numbers of a pion \cite%
{Voloshin:2018vym}. In accordance with this picture $Z_{c}$ and $%
Z_{c}^{-}(4200)$ are related by the charm quark spin symmetry which suggests
certain relations between their properties and decay channels. The possible
decays of a scalar and a vector tetraquark\ $[cd][\overline{c}\overline{u}]$
were analyzed also in Ref.\ \cite{Zhao:2018xrd}.

In the present work we treat $Z_{c}$ as the scalar diquark-antidiquark state
$[cd][\overline{c}\overline{u}]$, since it was observed in the process $%
Z_{c}\rightarrow \eta _{c}(1S)\pi ^{-}$. In fact, for the scalar $Z_{c}$
this decay is the dominant $S$-wave channel, whereas for the vector
tetraquark $Z_{c}$ it turns $P$-wave decay. We are going to calculate the
spectroscopic parameters of the tetraquark $Z_{c}$, i.e., its mass and
coupling by means of the two-point QCD sum rule method. The QCD sum rule
method is the powerful nonperturbative approach to investigate the
conventional hadrons and calculate their parameters \cite%
{Shifman:1978bx,Shifman:1978by}. But it can be successfully applied for
studying multiquark systems as well. To get reliable predictions for the
quantities of concern, in the sum rule computations we take into account the
quark, gluon, and mixed vacuum condensates up to dimension ten.

The next problem to be considered in this work is investigating decays of
the resonance $Z_{c}$ and evaluating its total width. It is known, that
strong and semileptonic decays of various tetraquark candidates provide
valuable information on their internal structure and dynamical features. In
the framework of the QCD sum rule approach relevant problems were subject of
rather intensive studies \cite%
{Navarra:2006nd,Dias:2013xfa,Agaev:2016dev,Agaev:2016ijz,Agaev:2016dsg,Agaev:2017uky,Agaev:2017foq, Wang:2017lot,Sundu:2018uyi,Sundu:2018toi,Agaev:2018vag,Agaev:2018khe,Wang:2018qpe,Wang:2019iaa}%
. The dominant strong decay of the resonance $Z_{c}$ seems is the channel $%
Z_{c}\rightarrow \eta _{c}(1S)\pi ^{-}$. But its $S$-wave hidden-charm $\eta
_{c}(2S)\pi ^{-}$, $J/\psi \rho ^{-}$ and open-charm $D^{0}D^{-}$ and $%
D^{\ast 0}D^{\ast -}$decays are also kinematically allowed modes \cite%
{Zhao:2018xrd}.

We calculate the partial width of the dominant $S$-wave processes and use
obtained results to evaluate the total width of the tetraquark $Z_{c}$. The
decays $Z_{c}\rightarrow \eta _{c}(1S)\pi ^{-}$ , $\eta _{c}(2S)\pi ^{-}$,
and $D^{0}D^{-}$ are explored by applying the QCD three-point sum rule
method. The quantities extracted from the sum rules are the strong couplings
$g_{Z_{c}\eta _{c1}\pi }$, $g_{Z_{c}\eta _{c2}\pi }$, and $g_{Z_{c}DD}$ that
correspond to the vertices $Z_{c}\eta _{c}(1S)\pi ^{-}$, $Z_{c}\eta
_{c}(2S)\pi ^{-}$, and $Z_{c}D^{0}D^{-}$, respectively. The coupling $%
g_{Z_{c}J/\psi \rho }$, which describes the strong vertex $Z_{c}J/\psi \rho
^{-}$, is found by means of the QCD light-cone sum rule (LCSR) method and
technical tools of the soft-meson approximation \cite%
{Balitsky:1989ry,Belyaev:1994zk}. For analysis of the tetraquarks this
method and approximation was adapted in Ref.\ \cite{Agaev:2016dev}, and
applied to study their numerous strong decay channels. Alongside the mass
and coupling of the state $Z_{c}$ the strong couplings provide an important
information to determine the width of the decays under analysis.

This work is organized in the following manner: In Sec.\ \ref{sec:Mass} we
calculate the mass $m$ and coupling $f$ of the scalar resonance $Z_{c}$ by
employing the two-point sum rule method and including into analysis the
quark, gluon, and mixed condensates up to dimension ten. The obtained
predictions for these parameters are applied in Sec.\ \ref{sec:Decays1} to
evaluate the partial widths of the decays $Z_{c}\rightarrow \eta _{c}(1S)\pi
^{-}$ and $\eta _{c}(2S)\pi ^{-}$. The decay $Z_{c}\rightarrow D^{0}D^{-}$
is considered in Sec. \ref{sec:Decay1a}, whereas Section \ref{sec:Decay2} is
devoted to analysis of the decay $Z_{c}\rightarrow J/\psi \rho ^{-}$. In
Sec. \ref{sec:Decay2} we also give our estimate for the total width of the
resonance $Z_{c}$. The Sec.\ \ref{sec:Conc} contains the analysis of
obtained results and our concluding notes. In the Appendix we write down
explicit expressions of the heavy and light quark propagators, as well as
the two-point spectral density used in the mass and coupling calculations.


\section{Mass and coupling of the scalar tetraquark $Z_{c}$}

\label{sec:Mass}
The scalar resonance $Z_{c}$ can be composed of the scalar diquark $\epsilon
^{ijk}[c_{j}^{T}C\gamma _{5}d_{k}]$ in the color antitriplet and flavor
antisymmetric state and the scalar antidiquark $\epsilon ^{imn}[\overline{c}%
_{m}\gamma _{5}C\overline{u}_{n}^{T}]$ in the color triplet state. These
diquarks are most attractive ones, and four-quark mesons composed of them
should be lighter and more stable than bound states of diquarks with other
quantum numbers \cite{Jaffe:2004ph}. The scalar diquarks were used as
building blocks to construct various hidden-charm and -bottom tetraquark
states and study their properties (see, for example Refs.\ \cite%
{Chen:2007xr,Wang:2009bd,Wang:2015gxa} ). In the present work for the
resonance $Z_{c}$ we choose namely this favorable structure.

To calculate the mass $m$ and coupling $f$ of the resonance $Z_{c}$ using
the QCD sum rule method, we start from the two-point correlation function
\begin{equation}
\Pi (p)=i\int d^{4}xe^{ip\cdot x}\langle 0|\mathcal{T}\{J(x)J^{\dagger
}(0)\}|0\rangle ,  \label{eq:CorrF1}
\end{equation}%
where $J(x)$ is the interpolating current for the tetraquark $Z_{c}$. In
accordance with our assumption on the structure of $Z_{c}$ the interpolating
current $J(x)$ has the following form
\begin{equation}
J(x)=\epsilon \tilde{\epsilon}\left[ c_{j}^{T}(x)C\gamma _{5}d_{k}(x)\right] %
\left[ \overline{c}_{m}(x)\gamma _{5}C\overline{u}_{n}^{T}(x)\right] ,
\label{eq:Curr}
\end{equation}%
Here we employ the notations $\epsilon =\epsilon ^{ijk}$ and $\tilde{\epsilon%
}=\epsilon ^{imn}$, where $i,j,k,m$ and $n$ are color indices, and $C$ is
the charge-conjugation operator.

To derive the sum rules for the mass $m$ and coupling $f$ of the
ground-state tetraquark $Z_{c}$ we adopt the "ground-state + continuum"
approximation, and calculate the physical or phenomenological side of the
sum rule. For these purposes, we insert into the correlation function a full
set of relevant states and carry out in Eq.\ (\ref{eq:CorrF1}) the
integration over $x$, and get
\begin{equation}
\Pi ^{\mathrm{Phys}}(p)=\frac{\langle 0|J|Z_{c}(p)\rangle \langle
Z_{c}(p)|J^{\dagger }|0\rangle }{m^{2}-p^{2}}+\ldots  \label{eq:Phys1}
\end{equation}%
Here we separate the ground-state contribution to $\Pi ^{\mathrm{Phys}}(p)$
from effects of the higher resonances and continuum states, which are
denoted there by the dots. In the calculations we assume that the
phenomenological side $\Pi ^{\mathrm{Phys}}(p)$ can be approximated by a
single pole term. In the case of the multiquark systems the physical side,
however, receives contribution also from two-meson reducible terms \cite%
{Kondo:2004cr,Lee:2004xk}. In other words, the interpolating current $J(x)$
interacts with the two-meson continuum, which generates the finite width $%
\Gamma (p^{2})$ of the tetraquark and results in the modification \cite%
{Wang:2015nwa}
\begin{equation}
\frac{1}{m^{2}-p^{2}}\rightarrow \frac{1}{m^{2}-p^{2}-i\sqrt{p^{2}}\Gamma
(p^{2})}.  \label{eq:Modification}
\end{equation}%
The two-meson continuum effects can be properly taken into account by
rescaling the coupling $f$, whereas the mass of the tetraquark $m$ preserves
its initial value. But these effects are numerically small, therefore in the
phenomenological side of the sum rule we use the zero-width single-pole
approximation and check afterwards its self-consistency.

Calculation of $\Pi ^{\mathrm{Phys}}(p)$ can be finished by introducing the
matrix element of the scalar tetraquark
\begin{equation}
\langle 0|J|Z_{c}\rangle =fm.  \label{eq:ME1}
\end{equation}%
As a result, we find%
\begin{equation}
\Pi ^{\mathrm{Phys}}(p)=\frac{f^{2}m^{2}}{m^{2}-p^{2}}+\ldots .
\label{eq:Phys2}
\end{equation}%
Because $\Pi ^{\mathrm{Phys}}(p)$ has trivial Lorentz structure proportional
to $I$, corresponding invariant amplitude $\Pi ^{\mathrm{Phys}}(p^{2})$ is
equal to the function given by Eq.\ (\ref{eq:Phys2}).

At the next step one has to find the correlation function $\Pi (p)$ using
the perturbative QCD and express it through the quark propagators, and, as a
result, in terms of the vacuum expectation values of various quark, gluon
and mixed operators as nonperturbative effects. To this end, we use the
interpolating current $J(x)$, contract the relevant heavy and light quark
operators in Eq.\ (\ref{eq:CorrF1}) to generate propagators, and obtain
\begin{eqnarray}
\Pi ^{\mathrm{OPE}}(p) &=&i\int d^{4}xe^{ip\cdot x}\epsilon \tilde{\epsilon}%
\epsilon ^{\prime }\tilde{\epsilon}^{\prime }\mathrm{Tr}\left[ \gamma _{5}%
\widetilde{S}_{c}^{jj^{\prime }}(x)\gamma _{5}S_{d}^{kk^{\prime }}(x)\right]
\notag \\
&&\times \mathrm{Tr}\left[ \gamma _{5}\widetilde{S}_{u}^{n^{\prime
}n}(-x)\gamma _{5}S_{c}^{m^{\prime }m}(-x)\right] .  \label{eq:OPE1}
\end{eqnarray}%
Here $S_{c}(x)$ and $S_{u(d)}(x)$ are the heavy $c$- and light $u(d)$-quark
propagators, respectively. These propagators contain both the perturbative
and nonperturbative components: their explicit expressions are presented in
the Appendix. In Eq.\ (\ref{eq:OPE1}) we also utilized the shorthand
notation
\begin{equation}
\widetilde{S}_{c(u)}(x)=CS_{c(u)}^{T}(x)C.  \label{eq:Notation}
\end{equation}

To extract the required sum rules for $m$ and $f$ one must equate $\Pi ^{%
\mathrm{Phys}}(p^{2})$ to the similar amplitude $\Pi ^{\mathrm{OPE}}(p^{2})$%
, apply the Borel transformation to both sides of the obtained equality to
suppress contributions of the higher resonances and, finally, perform the
continuum subtraction in accordance with the assumption on the quark-hadron
duality: These manipulations lead to the equality that can be used to get
the sum rules. The second equality, which is required for these purposes,
can be obtained from the first expression by applying on it by the operator $%
d/d(-1/M^{2})$. Then, for the mass of the tetraquark $Z_{c}$ we get the sum
rule
\begin{equation}
m^{2}=\frac{\int_{4m_{c}^{2}}^{s_{0}}dss\rho ^{\mathrm{OPE}}(s)e^{-s/M^{2}}}{%
\int_{4m_{c}^{2}}^{s_{0}}ds\rho ^{\mathrm{OPE}}(s)e^{-s/M^{2}}}.
\label{eq:Mass}
\end{equation}%
The sum rule for the coupling $f$ reads%
\begin{equation}
f^{2}=\frac{1}{m^{2}}\int_{4m_{c}^{2}}^{s_{0}}ds\rho ^{\mathrm{OPE}%
}(s)e^{(m^{2}-s)/M^{2}}.  \label{eq:Coupl}
\end{equation}%
where $M^{2}$ and $s_{0}$ are the Borel and continuum threshold parameters,
respectively. In Eqs.\ (\ref{eq:Mass}) and (\ref{eq:Coupl}) $\rho ^{\mathrm{%
OPE}}(s)$ is the two-point spectral density, which is proportional to the
imaginary part of the correlation function $\Pi ^{\mathrm{OPE}}(p).$ The
explicit expression of $\rho ^{\mathrm{OPE}}(s)$ is presented in the
Appendix.

We use the obtained sum rules to compute the mass $m$ and coupling $f$ of
the tetraquark $Z_{c}$. They contain numerous parameters, some of which,
such as the vacuum condensates, the mass of the $c$-quark, are universal
quantities and do not depend on the problem under discussion. In
computations we utilize the following values for the quark, gluon and mixed
condensates:
\begin{eqnarray}
&&\langle \bar{q}q\rangle =-(0.24\pm 0.01)^{3}\ \mathrm{GeV}^{3},\   \notag
\\
&&m_{0}^{2}=(0.8\pm 0.1)\ \mathrm{GeV}^{2},\ \langle \overline{q}g_{s}\sigma
Gq\rangle =m_{0}^{2}\langle \overline{q}q\rangle ,  \notag \\
&&\langle \frac{\alpha _{s}G^{2}}{\pi }\rangle =(0.012\pm 0.004)\,\mathrm{GeV%
}^{4},  \notag \\
&&\langle g_{s}^{3}G^{3}\rangle =(0.57\pm 0.29)~\mathrm{GeV}^{6}.
\label{eq:Parameters}
\end{eqnarray}%
The mass of the $c$-quark is taken equal to $m_{c}=1.275_{-0.035}^{+0.025}\
\mathrm{GeV}$.

The Borel parameter $M^{2}$ and continuum threshold $s_{0}$ are the
auxiliary parameters and should be chosen in accordance with standard
constraints accepted in the sum rule computations. The Borel parameter can
be varied within the limits $[M_{\mathrm{min}}^{2},\ M_{\mathrm{max}}^{2}]$
which have to obey the following conditions: At $M_{\mathrm{max}}^{2}$ the
pole contribution ($\mathrm{PC)}$ defined as the ratio
\begin{equation}
\mathrm{PC}=\frac{\Pi (M_{\mathrm{max}}^{2},\ s_{0})}{\Pi (M_{\mathrm{max}%
}^{2},\ \infty )},  \label{eq:PC}
\end{equation}%
should be larger than some fixed number. Let us note that $\Pi (M^{2},\
s_{0})$ in Eq.\ (\ref{eq:PC}) is the Borel transformed and subtracted
invariant amplitude $\Pi ^{\mathrm{OPE}}(p^{2})$. In the sum rule
calculations involving the tetraquarks the minimal value of $\mathrm{PC}$
varies between $0.15-0.2$. In the present work we choose $\mathrm{PC}>0.15$.
The minimal value of the Borel parameter $M_{\mathrm{min}}^{2}$ is fixed
from convergence of the sum rules: in other words, at $M_{\mathrm{min}}^{2}$
contribution of the last term (or a sum of last few terms) to $\Pi (M^{2},\
s_{0})$ cannot exceed $0.05$ part of the whole result
\begin{equation}
R(M_{\mathrm{min}}^{2})=\frac{\Pi ^{\mathrm{DimN}}(M_{\mathrm{min}}^{2},\
s_{0})}{\Pi (M_{\mathrm{min}}^{2},\ s_{0})}<0.05.  \label{eq:Convergence}
\end{equation}%
The ratio $R(M_{\mathrm{min}}^{2})$ quantifies the convergence of the OPE
and will be used for the numerical analysis. The last restriction on the
lower limit $M_{\mathrm{min}}^{2}$ is the prevalence of the perturbative
contribution over the nonperturbative one.

The mass $m$ and coupling $f$ should not depend on the parameters $M^{2}$
and $s_{0}$. But in real calculations, these quantities are sensitive to the
choice of $M^{2}$ and $s_{0}$. Therefore, the parameters $M^{2}$ and $s_{0}$
should also be determined in such a way that to minimize the dependence of $%
m $ and $f$ on them. The analysis allows us to fix the working windows for
the parameters $M^{2}$ and $s_{0}$
\begin{equation}
M^{2}\in \lbrack 4,\ 6]\ \mathrm{GeV}^{2},\ s_{0}\in \lbrack 19,\ 21]\
\mathrm{GeV}^{2},  \label{eq:Wind}
\end{equation}%
which obey all aforementioned restrictions. Thus, at $M^{2}=6~\mathrm{GeV}%
^{2}$ the pole contribution equals to $0.19$, and within the region $%
M^{2}\in \lbrack 4,~6]~\mathrm{GeV}^{2}$ it changes from $0.54$ till $0.19$.
To find the lower bound of the Borel parameter from Eq.\ (\ref%
{eq:Convergence}) we use the last three terms in the expansion, i.e. $%
\mathrm{DimN}=\mathrm{Dim(8+9+10)}$ [we remind that $\Pi ^{\mathrm{Dim10}}=0$%
]. Then at $M^{2}=4~\mathrm{GeV}^{2}$ the ratio $R$ becomes equal to $R(4~%
\mathrm{GeV}^{2})=0.02$ which ensures the convergence of the sum rules. At $%
M^{2}=4~\mathrm{GeV}^{2}$ the perturbative contribution amounts to $83\%$ of
the full result exceeding considerably the nonperturbative terms.

As it has been noted above, there are residual dependence of $m$ and $f$ on
the parameters $M^{2}$ and $s_{0}$. In Figs.\ \ref{fig:Mass1} and \ref%
{fig:Coupl1} we plot the mass and coupling of the tetraquark $Z_{c}$ as
functions of the parameters $M^{2}$ and $s_{0}$. It is seen that both the $m$
and $f$ depend on $M^{2}$ and $s_{0}$ which generates essential part of the
theoretical uncertainties inherent to the sum rule computations. For the
mass $m$ these uncertainties are small which has a simple explanation: The
sum rule for the mass (\ref{eq:Mass}) is equal to the ratio of integrals
over the functions $s\rho ^{\mathrm{OPE}}(s)$ and $\rho ^{\mathrm{OPE}}(s)$,
which reduces effects due to variation of $M^{2}$ and $s_{0}$. The coupling $%
f$ depends on the integral over the spectral density $\rho ^{\mathrm{OPE}%
}(s) $, and therefore its variations are sizeable. In the case under
analysis, theoretical errors for $m$ and $f$ generated by uncertainties of
various parameters including $M^{2}$ and $s_{0}$ ones equal to $\pm 3.7\%$
and $\pm 21\%$ of the corresponding central values, respectively.

Our analysis leads for the mass and coupling of the tetraquark $Z_{c}$ to
the following results:
\begin{eqnarray}
m &=&(4080~\pm 150)~\mathrm{MeV},  \notag \\
f &=&(0.58\pm 0.12)\cdot 10^{-2}\ \mathrm{GeV}^{4}.  \label{eq:CMass1}
\end{eqnarray}%
The mass of the resonance $Z_{c}$ modeled as the scalar diquark-antidiquark
state is in excellent agreement with the data of the LHCb Collaboration.

The scalar tetraquark $cu\overline{c}\overline{d}$ with $C\gamma _{5}\otimes
\gamma _{5}C$ structure was studied also in Ref.\ \cite{Wang:2017lbl}. The
prediction for the mass of this four-quark meson $m=(3860~\pm 90)~\mathrm{MeV%
}$ allowed the author to interpret it as the resonance $X^{\ast }(3860)$
(actually, its charged partner) observed recently by the Belle Collaboration
\cite{Chilikin:2017evr}. This charmoniumlike state was seen in the process $%
e^{+}e^{-}\rightarrow J/\psi D\overline{D}$, where $D$ refers to either $%
D^{0}$ or $D^{+}$ meson, and was considered there as a $\chi _{c0}(2P)$
candidate. Comparing our result and that of Ref.\ \cite{Wang:2017lbl}, one
can see the existence of an overlapping region between them, nevertheless
the difference between the central values $200\ ~\mathrm{MeV}$ is sizable.
This discrepancy is presumably connected with working regions for $M^{2}$
and $s_{0}$, and also with values of the vacuum condensates (fixed or
evolved) used in numerical computations.

\begin{widetext}

\begin{figure}[h!]
\begin{center}
\includegraphics[totalheight=6cm,width=8cm]{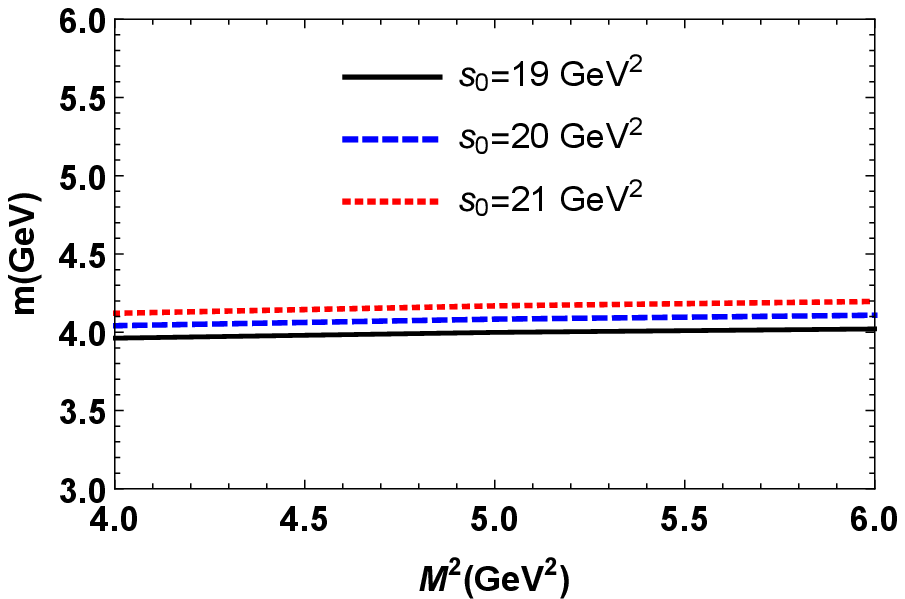}\,\, %
\includegraphics[totalheight=6cm,width=8cm]{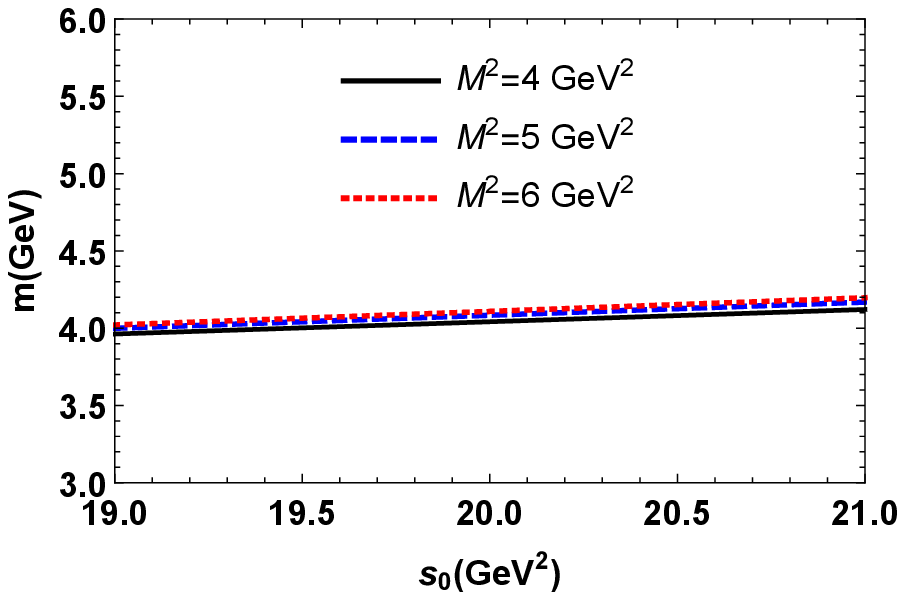}
\end{center}
\caption{ The mass of the state $Z_c^{-}(4100)$ as a function of the Borel parameter
$M^2$ at fixed $s_0$ (left panel), and as a function of the continuum threshold
$s_0$ at fixed $M^2$ (right panel).}
\label{fig:Mass1}
\end{figure}
\begin{figure}[h!]
\begin{center}
\includegraphics[totalheight=6cm,width=8cm]{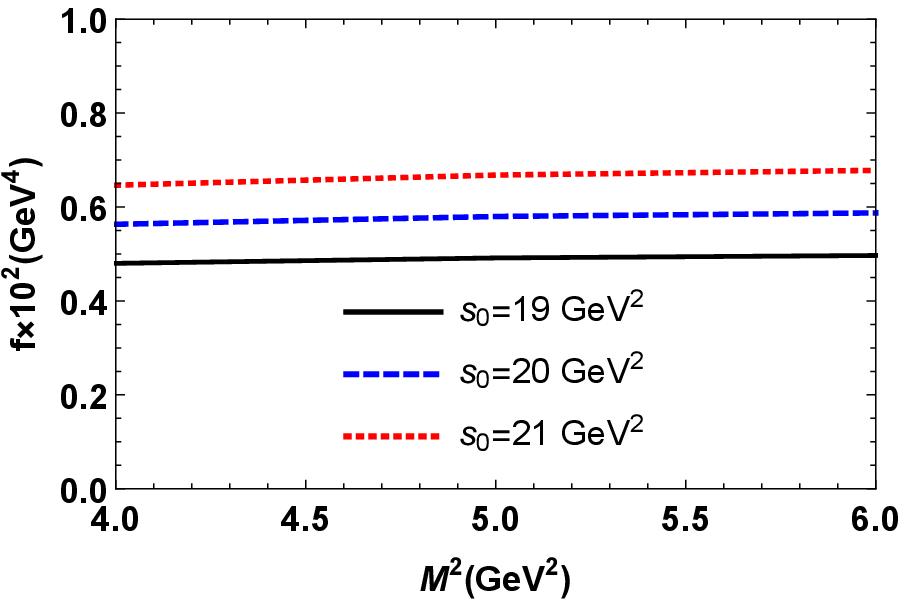}\,\, %
\includegraphics[totalheight=6cm,width=8cm]{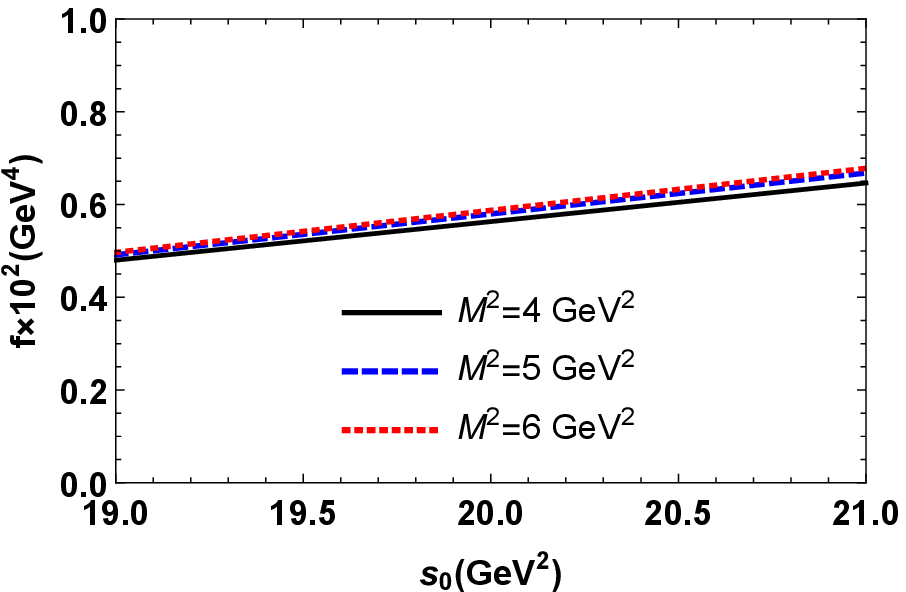}
\end{center}
\caption{ The same as in Fig. 1, but for the coupling $f$ of the resonance $Z_c^{-}(4100)$.}
\label{fig:Coupl1}
\end{figure}

\end{widetext}


\section{Decays $Z_{c}\rightarrow \protect\eta _{c}(1S)\protect\pi ^{-}$ and
$Z_{c}\rightarrow \protect\eta _{c}(2S)\protect\pi ^{-}$}

\label{sec:Decays1}

The $S$-wave decays of the resonance $Z_{c}$ can be divided into two
subclasses: The decays to two pseudoscalar and two vector mesons,
respectively. The processes $Z_{c}\rightarrow \eta _{c}(1S)\pi ^{-}$ and $%
Z_{c}\rightarrow \eta _{c}(2S)\pi ^{-}$ belong to the first subclass of
decays. The final stages of these decays contain the ground-state and first
radially excited $\eta _{c}$ mesons, therefore in the QCD sum rule approach
they should investigated in a correlated form. An appropriate way to deal
with decays $Z_{c}\rightarrow \eta _{c}(1S)\pi ^{-}$ and $Z_{c}\rightarrow
\eta _{c}(2S)\pi ^{-}$ is the QCD three-point sum rule method. Indeed,
because we are going to explore the form factors $g_{Z_{c}\eta _{ci}\pi
}(q^{2})$ for the off-shell pion the double Borel transformation will be
carried out in the $Z_{c}$ and $\eta _{c}$ channels, i.e. over momenta of
these particles. This transformation applied to the phenomenological side of
the relevant three-point sum rules suppresses contributions of the higher
resonances in these two channels eliminating, at the same time, terms
associated with the pole-continuum transitions \cite%
{Belyaev:1994zk,Ioffe:1983ju}. The elimination of these terms is important
for joint analysis of the form factors $g_{Z_{c}\eta _{ci}\pi }(q^{2})$,
because one does not need to apply an additional operator to remove them
from the phenomenological side of the sum rules. Nevertheless, there may
still exist in the pion channel terms corresponding to excited states of the
pion which emerge as contaminations [for the $NN\pi $ vertex, see
discussions in Refs.\ \cite{Meissner:1995ra,Maltman:1997jb}]. To reduce the
uncertainties in evaluation of the strong couplings at the vertices and
smooth problems with extrapolation of the form factors to the mass-shell, it
is possible to fix the pion on the mass-shell and treat one of the remaining
heavy states ($Z_{c}$ or $\eta _{c}$) as the off-shell particle. This trick
was used numerously to study the conventional heavy-heavy-light mesons'
couplings \cite{Bracco:2006xf,Cerqueira:2015vva}. Form factors obtained by
treating a light or one of heavy mesons off-shell may differ from each other
considerably, but after extrapolating to the corresponding mass-shells lead
to the same or slightly different strong couplings.

In the framework of the three-point sum rule approach a more detailed
representation for the phenomenological side was used in Refs.\ \cite%
{Wang:2017lot,Wang:2018qpe,Wang:2019iaa}. This technique generates
additional terms in the sum rules and introduces into analysis new free
parameters, which should be chosen to obtain stable sum rules with
variations of the Borel parameters. In the present work, to calculate $%
g_{Z_{c}\eta _{c1}\pi }(q^{2})$ and $g_{Z_{c}\eta _{c2}\pi }(q^{2})$ we
apply the standard three-point sum rule method and choose the pion an
off-shell particle. We use this method to study the decay $Z_{c}\rightarrow
D^{0}D^{-}$ as well.

The process $Z_{c}\rightarrow $ $J/\psi \rho ^{-}$ belongs to the second
subclass of $Z_{c}$ decays; it is a decay to two vector mesons. We
investigate this mode by means of the QCD light-cone sum rule method and
soft-meson approximation. The sum rule on the light-cone allows one to find
the strong coupling by avoiding extrapolating procedures and express $%
g_{Z_{c}J/\psi \rho }$ not only in terms of the vacuum condensates, but also
using the $\rho $-meson local matrix elements. As for unsuppressed
pole-continuum effects that after a single Borel transformation survive in
this approach, they can be eliminated by means of well-known prescriptions
\cite{Ioffe:1983ju}.

To determine the partial widths of the decays $Z_{c}\rightarrow \eta
_{c}(1S)\pi ^{-}$ and $Z_{c}\rightarrow \eta _{c}(2S)\pi ^{-}$ one needs to
calculate the strong couplings $g_{Z_{c}\eta _{c1}\pi }$ and $g_{Z_{c}\eta
_{c2}\pi }$ which can be extracted from the three-point correlation function%
\begin{eqnarray}
\Pi (p,p^{\prime }) &=&i^{2}\int d^{4}xd^{4}ye^{-ip\cdot x}e^{ip^{\prime
}\cdot y}  \notag \\
&&\times \langle 0|\mathcal{T}\{J^{\eta }(y)J^{\pi }(0)J^{\dagger
}(x)\}|0\rangle ,  \label{eq:CorrF1A}
\end{eqnarray}%
where
\begin{equation}
J^{\eta }(y)=\overline{c}_{a}(y)i\gamma _{5}c_{a}(y),\ J^{\pi }(0)=\overline{%
u}_{b}(0)i\gamma _{5}d_{b}(0)  \label{eq:Curr1A}
\end{equation}%
are the interpolating currents for the pseudoscalar mesons $\eta _{c}$ and $%
\pi ^{-}$, respectively. The $J(x)$ is the interpolating current for the
resonance $Z_{c}$ and has been introduced above in Eq.\ (\ref{eq:Curr}).

In terms of the physical parameters of the tetraquark and mesons the
correlation function $\Pi (p,p^{\prime })$ takes the form
\begin{eqnarray}
&&\Pi ^{\mathrm{Phys}}(p,p^{\prime })=\sum_{i=1}^{2}\frac{\langle 0|J^{\eta
}|\eta _{ci}\left( p^{\prime }\right) \rangle }{p^{\prime 2}-m_{i}^{2}}\frac{%
\langle 0|J^{\pi }|\pi \left( q\right) \rangle }{q^{2}-m_{\pi }^{2}}  \notag
\\
&&\times \frac{\langle \eta _{ci}\left( p^{\prime }\right) \pi
(q)|Z_{c}(p)\rangle \langle Z_{c}(p)|J^{\dagger }|0\rangle }{p^{2}-m^{2}}%
+\ldots ,  \label{eq:PhysD1}
\end{eqnarray}%
where $m_{\pi }$ is the mass of the pion, and $m_{i}=m_{1}$, $m_{2}$ are
masses of the mesons $\eta _{c}(1S)$ and $\eta _{c}(2S)$, respectively. The
four-momenta of the particles are evident from (\ref{eq:PhysD1}). Here by
the dots we denote contribution of the higher resonances and continuum
states.

To continue we introduce the matrix elements
\begin{equation}
\langle 0|J^{\eta }|\eta _{ci}\left( p^{\prime }\right) \rangle =\frac{%
f_{i}m_{i}^{2}}{2m_{c}},\,  \label{eq:ME2}
\end{equation}%
where $f_{1}$ and $f_{2}$ are the decay constants of the mesons $\eta
_{c}(1S)$ and $\eta _{c}(2S)$, respectively. The relevant matrix element of
the pion is well known and has the form%
\begin{equation}
\langle 0|J^{\pi }|\pi \left( q\right) \rangle =\mu _{\pi }f_{\pi },\ \mu
_{\pi }=-\frac{2\langle \overline{q}q\rangle }{f_{\pi }^{2}},\
\label{eq:MEpion}
\end{equation}%
where $f_{\pi }$ is the decay constant of the pion, and $\langle \overline{q}%
q\rangle $ is the quark condensate. Additionally, the matrix elements of the
vertices $Z_{c}\eta _{c}(1S)\pi ^{-}$ and $\ Z_{c}\eta _{c}(2S)\pi ^{-}$ are
required. To this end, we use
\begin{equation}
\langle \eta _{ci}\left( p^{\prime }\right) \pi (q)|Z_{c}(p)\rangle
=g_{Z_{c}\eta _{ci}\pi }(p\cdot p^{\prime }).  \label{eq:ME3}
\end{equation}%
Here, the strong coupling $g_{Z_{c}\eta _{c1}\pi }$ corresponds to the
vertex $Z_{c}\eta _{c}(1S)\pi ^{-}$, whereas $g_{Z_{c}\eta _{c2}\pi }$
describes $Z_{c}\eta _{c}(2S)\pi ^{-}$; namely these couplings have to be
determined from the sum rules.

Employing Eqs.\ (\ref{eq:ME2}), (\ref{eq:MEpion}) and (\ref{eq:ME3}) for $%
\Pi ^{\mathrm{Phys}}(p,p^{\prime })$ we get the simple expression:
\begin{eqnarray}
&&\Pi ^{\mathrm{Phys}}(p,p^{\prime })=\sum_{i=1}^{2}\frac{g_{Z_{c}\eta
_{ci}\pi }m_{i}^{2}f_{i}mf}{2m_{c}(p^{\prime 2}-m_{i}^{2})\left(
p^{2}-m^{2}\right) }  \notag \\
&&\times \frac{\mu _{\pi }f_{\pi }}{q^{2}-m_{\pi }^{2}}(p\cdot p^{\prime
})+\dots .  \label{eq:Phys4}
\end{eqnarray}%
The Lorentz structure of the $\Pi ^{\mathrm{Phys}}(p,p^{\prime })$ is
proportional to $I$ therefore the invariant amplitude $\Pi ^{\mathrm{Phys}%
}(p^{2},p^{\prime 2},q^{2})$ is given by the sum of two terms from Eq.\ (\ref%
{eq:Phys4}). The double Borel transformation of $\Pi ^{\mathrm{Phys}%
}(p^{2},p^{\prime 2},q^{2})$ over the variables $p^{2}$ and $p^{\prime 2}$
with the parameters $M_{1}^{2}$ and $M_{2}^{2}$ forms one of sides in the
sum rule equality.

The QCD side of the sum rule, i.e. the expression of the correlation
function in terms of the quark propagators reads%
\begin{eqnarray}
&&\Pi ^{\mathrm{OPE}}(p,p^{\prime })=i^{2}\int d^{4}xd^{4}ye^{-ip\cdot
x}e^{ip^{\prime }\cdot y}\epsilon ^{ijk}\epsilon ^{imn}  \notag \\
&&\times \mathrm{Tr}\left[ \gamma _{5}S_{c}^{aj}(y-x)\gamma _{5}\widetilde{S}%
_{d}^{bk}(-x)\gamma _{5}\widetilde{S}_{u}^{nb}(x)\gamma _{5}S_{c}^{ma}(x-y)%
\right] .  \notag \\
&&  \label{eq:OPEDec}
\end{eqnarray}%
The Borel transformation $\mathcal{B}\Pi ^{\mathrm{OPE}}(p^{2},p^{\prime
2},q^{2})$, where $\Pi ^{\mathrm{OPE}}(p^{2},p^{\prime 2},q^{2})$ is the
invariant amplitude that corresponds to the structure $\sim I$ in $\Pi ^{%
\mathrm{OPE}}(p,p^{\prime })$ constitutes the second component of the sum
rule. Equating $\mathcal{B}\Pi ^{\mathrm{OPE}}(p^{2},p^{\prime 2},q^{2})$
and the double Borel transformation of $\Pi ^{\mathrm{Phys}}(p^{2},p^{\prime
2},q^{2})$ and performing continuum subtraction we find the sum rule for the
couplings $g_{Z_{c}\eta _{c1}\pi }$ and $g_{Z_{c}\eta _{c2}\pi }$.

The Borel transformed and subtracted amplitude $\Pi ^{\mathrm{OPE}%
}(p^{2},p^{\prime 2},q^{2})$ can be expressed in terms of the spectral
density $\rho _{\mathrm{D}}(s,s^{\prime },q^{2})$ which is proportional to
the imaginary part of $\Pi ^{\mathrm{OPE}}(p,p^{\prime })$
\begin{eqnarray}
&&\Pi (\mathbf{M}^{2},\mathbf{\ s}_{0},~q^{2})=\int_{4m_{c}^{2}}^{s_{0}}ds%
\int_{4m_{c}^{2}}^{s_{0}^{\prime }}ds^{\prime }\rho _{\mathrm{D}%
}(s,s^{\prime },q^{2})  \notag \\
&&\times e^{-s/M_{1}^{2}}e^{-s^{\prime }/M_{2}^{2}},  \label{eq:SCoupl}
\end{eqnarray}%
where $\mathbf{M}^{2}=(M_{1}^{2},\ M_{2}^{2})$ and $\mathbf{s}_{0}=(s_{0},\
s_{0}^{\prime })$ are the Borel and continuum threshold parameters,
respectively.

The obtained sum rule has to be used to determine the couplings $%
g_{Z_{c}\eta _{c1}\pi }$ and $g_{Z_{c}\eta _{c2}\pi }$. A possible way to
find them is to get the second sum rule from the first one by applying the
operators $d/d(-1/M_{1}^{2})$ and/or $d/d(-1/M_{2}^{2})$. But in the present
work we choose the alternative approach and use iteratively the master sum
rule to extract both $g_{Z_{c}\eta _{c1}\pi }$ and $g_{Z_{c}\eta _{c2}\pi }$%
. To this end, we fix the continuum threshold parameter $\sqrt{s_{0}^{\prime
}}$ which corresponds to the $\eta _{c}$ channel just below the mass of the
first radially excited state $\eta _{c}(2S)$. By this manipulation we
include $\eta _{c}(2S)$ into the continuum and obtain the sum rule for the
strong coupling of the ground-state meson $\eta _{c}(1S).$ The physical side
of the sum rule \ (\ref{eq:Phys4}) at this stage contains only the
ground-state term and depends on the coupling $g_{Z_{c}\eta _{c1}\pi }$.
This sum rule can be easily solved to evaluate the unknown parameter $%
g_{Z_{c}\eta _{c1}\pi }$
\begin{equation}
g_{Z_{c}\eta _{c1}\pi }(\mathbf{M}^{2},\mathbf{\ s}_{0}^{(1)},~q^{2})=\frac{%
\Pi (\mathbf{M}^{2},\mathbf{\ s}%
_{0}^{(1)},~q^{2})e^{m/M_{1}^{2}}e^{m_{1}^{2}/M_{2}^{2}}}{A_{1}},
\label{eq:StC1}
\end{equation}%
where
\begin{equation*}
A_{1}=\frac{mfm_{1}^{2}f_{1}\mu _{\pi }f_{\pi }}{4m_{c}(q^{2}-m_{\pi }^{2})}%
\left( m^{2}+m_{1}^{2}-q^{2}\right) ,
\end{equation*}%
and $\mathbf{\ s}_{0}^{(1)}=(s_{0},\ s_{0}^{\prime }\simeq m_{2}^{2}).$

At the next step we fix the continuum threshold $\sqrt{s_{0}^{\prime }}$ at $%
m_{2}+(0.5-0.8)\ \mathrm{GeV}$ and use the sum rule that now contains the
ground and first radially excited states. The QCD side of this sum rule is
given by the expression $\Pi (\mathbf{M}^{2},\mathbf{\ s}_{0}^{(2)},~q^{2})$
with $\mathbf{s}_{0}^{(2)}=(s_{0},\ s_{0}^{\prime }\simeq \lbrack
m_{2}+(0.5-0.8)]^{2}\ )$. By substituting the obtained expression for $%
g_{Z_{c}\eta _{c1}\pi }$ into this sum rule it is not difficult to evaluate
the second coupling $g_{Z_{c}\eta _{c2}\pi }$.

The couplings depend on the Borel and continuum threshold parameters and, at
the same time, are functions of $q^{2}$. In what follows we omit their
dependence on the parameters, replace $q^{2}=-Q^{2}$ and denote the obtained
couplings as $g_{Z_{c}\eta _{c1}\pi }(Q^{2})$ and $g_{Z_{c}\eta _{c2}\pi
}(Q^{2})$. For calculation of the decay width we need value of the strong
couplings at the pion's mass-shell, i.e. at $q^{2}=m_{_{\pi }}^{2}$, which
is not accessible for the sum rule calculations. The standard way to avoid
this problem is to introduce a fit functions $F_{1(2)}(Q^{2})$ that for the
momenta $Q^{2}>0$ leads to the same predictions as the sum rules, but can be
readily extrapolated to the region of $Q^{2}<0$. Let us emphasize that
values of the fit functions at the mass-shell are the strong couplings $%
g_{Z_{c}\eta _{c1}\pi }$ and $g_{Z_{c}\eta _{c2}\pi }$ to be utilized in
calculations.

Expressions for $g_{Z_{c}\eta _{c1}\pi }(Q^{2})$ and $g_{Z_{c}\eta _{c2}\pi
}(Q^{2})$ depend on various constants, such as the masses and decay
constants of the final-state mesons. The values of these parameters are
collected in Table \ref{tab:Param}. Additionally, there are parameters $%
\mathbf{M}^{2}$ and $\mathbf{s}_{0}$ which should also be fixed to carry out
numerical analysis. The requirements imposed on these auxiliary parameters
have been discussed above and are standard for all sum rule computations.
The regions for $M_{1}^{2}$ and $s_{0}$ which correspond to the tetraquark $%
Z_{c}$ coincide with the working windows for these parameters fixed in the
mass calculations $M_{1}^{2}\in \lbrack 4,\ 6]\ \mathrm{GeV}^{2},\ s_{0}\in
\lbrack 19,\ 21]\ \mathrm{GeV}^{2}$. The Borel and continuum threshold
parameters $M_{2}^{2},\ s_{0}^{\prime }$ in Eq.\ (\ref{eq:StC1}) are chosen
as
\begin{equation}
M_{2}^{2}\in \lbrack 3,\ 4]\ \mathrm{GeV}^{2},\ s_{0}^{\prime }=13\ \mathrm{%
GeV}^{2},  \label{eq:Wind2}
\end{equation}%
whereas in the sum rule for the second coupling $g_{Z_{c}\eta _{c2}\pi
}(Q^{2})$ we employ
\begin{equation}
M_{2}^{2}\in \lbrack 3,\ 4]\ \mathrm{GeV}^{2},\ s_{0}^{\prime }\in \lbrack
17,\ 19]\ \mathrm{GeV}^{2},  \label{eq:Wind3}
\end{equation}

As it has been emphasized above to evaluate the strong couplings at the
mass-shell $Q^{2}=-m_{\pi }^{2}$ we need to determine the fit functions. To
this end, we employ the following functions

\begin{equation}
F_{i}(Q^{2})=F_{0}^{i}\mathrm{\exp }\left[ c_{1i}\frac{Q^{2}}{m^{2}}%
+c_{2i}\left( \frac{Q^{2}}{m^{2}}\right) ^{2}\right] ,  \label{eq:FitF}
\end{equation}%
where $F_{0}^{i}$, $c_{1i}$ and $c_{2i}$ are fitting parameters. The
performed analysis allows us to find the parameters as $F_{0}^{1}=0.49\
\mathrm{GeV}^{-1}$, $c_{11}=27.64$ and $c_{12}=-34.66$. Another set reads $%
F_{0}^{2}=0.39\ \mathrm{GeV}^{-1}$, $c_{21}=28.13$ and $c_{22}=-35.24$.

At the mass-shell the strong couplings are equal to
\begin{eqnarray}
g_{Z_{c}\eta _{c1}\pi }(-m_{\pi }^{2}) &=&(0.47\pm 0.06)\ \mathrm{GeV}^{-1},
\notag \\
g_{Z_{c}\eta _{c2}\pi }(-m_{\pi }^{2}) &=&(0.38\pm 0.05)\ \mathrm{GeV}^{-1}.
\label{eq:MassSC}
\end{eqnarray}%
The widths of the decays $Z_{c}\rightarrow \eta _{c}(1S)\pi ^{-}$ and $%
Z_{c}\rightarrow \eta _{c}(2S)\pi ^{-}$ can be found by means of the formula%
\begin{eqnarray}
&&\Gamma \left[ Z_{c}\rightarrow \eta _{c}(\mathrm{I}S)\pi ^{-}\right] =%
\frac{g_{Z_{c}\eta _{ci}\pi }^{2}m_{i}^{2}}{8\pi }\lambda \left(
m,m_{i},m_{\pi }\right)  \notag \\
&&\times \left[ 1+\frac{\lambda ^{2}\left( m,m_{i},m_{\pi }\right) }{%
m_{i}^{2}}\right] ,\ \mathrm{I}\equiv i=1,2  \label{eq:DW1a}
\end{eqnarray}%
where%
\begin{equation*}
\lambda \left( a,b,c\right) =\frac{1}{2a}\sqrt{a^{4}+b^{4}+c^{4}-2\left(
a^{2}b^{2}+a^{2}c^{2}+b^{2}c^{2}\right) }.
\end{equation*}%
For the decay $Z_{c}\rightarrow \eta _{c}(1S)\pi ^{-}$
one has to set $g_{Z_{c}\eta _{ci}\pi }\rightarrow g_{Z_{c}\eta _{c1}\pi }$
and $m_{i}\rightarrow m_{1}$, whereas in the case of $Z_{c}\rightarrow \eta
_{c}(2S)\pi ^{-}$ quantities with subscript $2$ have to be used.

Using the strong couplings given by Eq.\ (\ref{eq:MassSC}) and Eq.\ (\ref%
{eq:DW1a}) it is not difficult to evaluate the partial widths of the decay
channels%
\begin{eqnarray}
\Gamma \left[ Z_{c}\rightarrow \eta _{c}(1S)\pi ^{-}\right] &=&(81\pm 17)\
\mathrm{MeV,}  \notag \\
\Gamma \left[ Z_{c}\rightarrow \eta _{c}(2S)\pi ^{-}\right] &=&(32\pm 7)\
\mathrm{MeV,}  \label{eq:DW1Numeric}
\end{eqnarray}%
which are main results of this section.
\begin{table}[tbp]
\begin{tabular}{|c|c|}
\hline\hline
Parameters & Values (in $\mathrm{MeV}$ units) \\ \hline\hline
$m_1=m[\eta_c(1S)]$ & $2983.9\pm 0.5$ \\
$f_1=f[\eta_c(1S)]$ & $404$ \\
$m_2=m[\eta_c(2S)]$ & $3637.6\pm 1.2$ \\
$f_2=f[\eta_c(2S)]$ & $331$ \\
$m_{J/\psi}$ & $3096.900\pm 0.006$ \\
$f_{J/\psi}$ & $411 \pm 7$ \\
$m_{\pi}$ & $139.57077 \pm 0.00018$ \\
$f_{\pi}$ & $131.5 $ \\
$m_{\rho}$ & $775.26\pm 0.25$ \\
$f_{\rho}$ & $216 \pm 3$ \\
$m_{D^{0}}$ & $1864.83\pm 0.05$ \\
$m_{D}$ & $1869.65\pm 0.05$ \\
$f_{D}=f_{D^{0}}$ & $211.9 \pm 1.1$ \\ \hline\hline
\end{tabular}%
\caption{Parameters of the mesons produced in the decays of the resonance $%
Z_{c}$.}
\label{tab:Param}
\end{table}


\section{Decay $Z_{c}\rightarrow D^{0}D^{-}$}

\label{sec:Decay1a}

This section is devoted to investigation of the process $Z_{c}\rightarrow
D^{0}D^{-}$, which is the $S$-wave decay to the two open-charm pseudoscalar
mesons. Our starting point is the three-point correlation function
\begin{eqnarray}
\widetilde{\Pi }(p,p^{\prime }) &=&i^{2}\int d^{4}xd^{4}ye^{-ip\cdot
x}e^{ip^{\prime }\cdot y}  \notag \\
&&\times \langle 0|\mathcal{T}\{J^{D}(y)J^{D^{0}}(0)J^{\dagger
}(x)\}|0\rangle ,  \label{eq:DCF1}
\end{eqnarray}%
where
\begin{equation}
\ J^{D}(0)=\overline{c}_{r}(y)i\gamma _{5}d_{r}(y),\ J^{D^{0}}(y)=\overline{u%
}_{s}(0)i\gamma _{5}c_{s}(0),  \label{eq:Dcurrs}
\end{equation}%
are the interpolating currents for the pseudoscalar mesons $D^{-}$ and $%
D^{0} $, respectively.

The correlation function $\Pi (p,p^{\prime })$ expressed using parameters of
the mesons $D^{0}$ and $D^{-}$ and tetraquark $Z_{c}$ has the form
\begin{eqnarray}
&&\widetilde{\Pi }^{\mathrm{Phys}}(p,p^{\prime })=\frac{\langle
0|J^{D}|D^{-}\left( p^{\prime }\right) \rangle }{p^{\prime 2}-m_{D}^{2}}%
\frac{\langle 0|J^{D^{0}}|D^{0}\left( q\right) \rangle }{q^{2}-m_{D^{0}}^{2}}
\notag \\
&&\times \frac{\langle D^{-}\left( p^{\prime }\right)
D^{0}(q)|Z_{c}(p)\rangle \langle Z_{c}(p)|J^{\dagger }|0\rangle }{p^{2}-m^{2}%
}+\ldots ,  \label{eq:DCF2}
\end{eqnarray}%
where $m_{D}$ and $m_{D^{0}}$ are masses of the mesons $D^{-}$ and $D^{0}$,
respectively. Contribution of the higher resonances and continuum states, as
usual, are shown by dots.

We continue by utilizing the matrix elements
\begin{eqnarray}
\langle 0|J^{D}|D^{-}\left( p^{\prime }\right) \rangle &=&\frac{%
f_{D}m_{D}^{2}}{m_{c}}  \notag \\
\langle 0|J^{D^{0}}|D^{0}\left( q\right) \rangle &=&\frac{%
f_{D^{0}}m_{D^{0}}^{2}}{m_{c}}  \label{eq:DME1}
\end{eqnarray}%
and the vertex%
\begin{equation}
\langle D^{-}\left( p^{\prime }\right) D^{0}(q)|Z_{c}(p)\rangle
=g_{Z_{c}DD}(p\cdot p^{\prime }).  \label{eq:DME2}
\end{equation}%
Simple manipulations lead to:%
\begin{eqnarray}
&&\widetilde{\Pi }^{\mathrm{Phys}}(p,p^{\prime })=\frac{%
f_{D^{0}}m_{D^{0}}^{2}f_{D}m_{D}^{2}}{m_{c}^{2}\left( p^{\prime
2}-m_{D}^{2}\right) \left( q^{2}-m_{D^{0}}^{2}\right) }  \notag \\
&&\times \frac{mf}{p^{2}-m^{2}}(p\cdot p^{\prime })+\dots .  \label{eq:DCF3}
\end{eqnarray}%
Because the Lorentz structure of $\ \widetilde{\Pi }^{\mathrm{Phys}%
}(p,p^{\prime })$ is trivial and $\sim I$ , the invariant amplitude $%
\widetilde{\Pi }^{\mathrm{Phys}}(p^{2},p^{\prime 2},q^{2})$ is given by the
function from Eq.\ (\ref{eq:DCF3}).

The same correlation function written down in terms of the quark propagators
is%
\begin{eqnarray}
&&\widetilde{\Pi }^{\mathrm{OPE}}(p,p^{\prime })=i^{2}\int
d^{4}xd^{4}ye^{-ip\cdot x}e^{ip^{\prime }\cdot y}\epsilon ^{ijk}\epsilon
^{imn}  \notag \\
&&\times \mathrm{Tr}\left[ \gamma _{5}S_{d}^{rk}(y-x)\gamma _{5}\widetilde{S}%
_{c}^{sj}(-x)\gamma _{5}\widetilde{S}_{u}^{ns}(x)\gamma _{5}S_{c}^{mr}(x-y)%
\right] .  \notag \\
&&  \label{eq:DCF4}
\end{eqnarray}%
The invariant amplitudes $\widetilde{\Pi }^{\mathrm{OPE}}(p^{2},p^{\prime
2},q^{2})$ and $\widetilde{\Pi }^{\mathrm{Phys}}(p^{2},p^{\prime 2},q^{2})$
equated to each other yield the required sum rule. Contributions of higher
resonances and continuum states can be suppressed by applying the double
Borel transformation, and subtracted in accordance with the quark-hadron
duality assumption.

The final sum rule for the strong coupling can be recast to the traditional
form
\begin{equation}
g_{Z_{c}DD}(\mathbf{M}^{2},\mathbf{\ s}_{0},~q^{2})=\frac{\widetilde{\Pi }(%
\mathbf{M}^{2},\mathbf{\ s}_{0},~q^{2})e^{m/M_{1}^{2}}e^{m_{D}^{2}/M_{2}^{2}}%
}{B},  \label{eq:DSR}
\end{equation}%
where
\begin{equation*}
B=\frac{f_{D}m_{D}^{2}f_{D^{0}}m_{D^{0}}^{2}mf}{%
2m_{c}^{2}(q^{2}-m_{D^{0}}^{2})}\left( m^{2}+m_{D}^{2}-q^{2}\right) .
\end{equation*}%
Here $\widetilde{\Pi }(\mathbf{M}^{2},\mathbf{\ s}_{0},~q^{2})$ is the
Borel-transformed and continuum-subtracted amplitude $\widetilde{\Pi }^{%
\mathrm{OPE}}(p^{2},p^{\prime 2},q^{2})$ given by analogous to Eq. \ (\ref%
{eq:DSR}) formula:%
\begin{eqnarray}
&&\widetilde{\Pi }(\mathbf{M}^{2},\mathbf{\ s}_{0},~q^{2})=%
\int_{4m_{c}^{2}}^{s_{0}}ds\int_{m_{c}^{2}}^{s_{0}^{\prime }}ds^{\prime }%
\widetilde{\rho }_{\mathrm{D}}(s,s^{\prime },q^{2})  \notag \\
&&\times e^{-s/M_{1}^{2}}e^{-s^{\prime }/M_{2}^{2}}.  \label{eq:DCF5}
\end{eqnarray}

The sum rule for the strong coupling $g_{Z_{c}DD}$ depends on vacuum
condensates, and contains also the masses and decay constants of the mesons $%
D^{0}$ and $D^{-}$, which are shown in Table \ref{tab:Param}. Constraints
imposed on the auxiliary parameters $\mathbf{M}^{2}$ and $\mathbf{s}_{0}$
are similar to ones discussed above and universal for all sum rules
computations.The parameters $M_{1}^{2}$ and $s_{0}$ coincide with the
working regions for these parameters fixed in the mass calculations (\ref%
{eq:Wind}). The Borel and continuum threshold parameters $M_{2}^{2},\
s_{0}^{\prime }$ in Eq.\ (\ref{eq:DCF5})
\begin{equation}
M_{2}^{2}\in \lbrack 3,\ 6]\ \mathrm{GeV}^{2},\ s_{0}^{\prime }\in \lbrack
7,\ 9]\ \mathrm{GeV}^{2},  \label{eq:Wind4}
\end{equation}
and ones from Eq.\ (\ref{eq:Wind}) lead to stable results for the form
factor $g_{Z_{c}DD}(\mathbf{M}^{2},\mathbf{\ s}_{0},~q^{2})$ at $q^{2}<0$.
In what follows we denote it $g_{Z_{c}DD}(Q^{2})$ omitting dependence on $%
\mathbf{(M}^{2},\mathbf{\ s}_{0})$ and introducing $q^{2}=-Q^{2}$.

\begin{figure}[h]
\includegraphics[width=8.8cm]{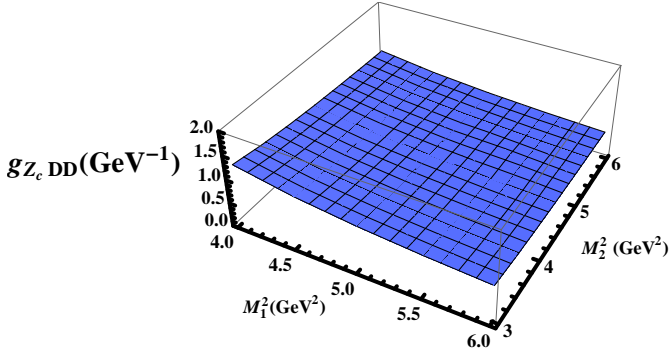}
\caption{The strong coupling $g_{Z_{c}DD}(Q^{2})$ as a function of the Borel
parameters $\mathbf{M}^{2}=(M_{1}^{2},\ M_{2}^{2})$ at the fixed $%
(s_{0},s_{0}^{\prime })=(20,8)\ \mathrm{GeV}^{2}$ and $Q^{2}=5~\mathrm{GeV}%
^{2}$.}
\label{fig:3Dplot}
\end{figure}

A sensitivity of  the strong coupling $g_{Z_{c}DD}(Q^{2})$ to the Borel
parameters is demonstrated in Fig.\ \ref{fig:3Dplot}, which   reveals its
residual dependence on $M_{1}^{2}$ and $M_{2}^{2}$. This dependence of $%
g_{Z_{c}DD}(Q^{2})$ as well as its variations generated by the continuum
threshold parameters are main sources of ambiguities in the sum rule
computations.

\begin{figure}[h]
\includegraphics[width=8.5cm]{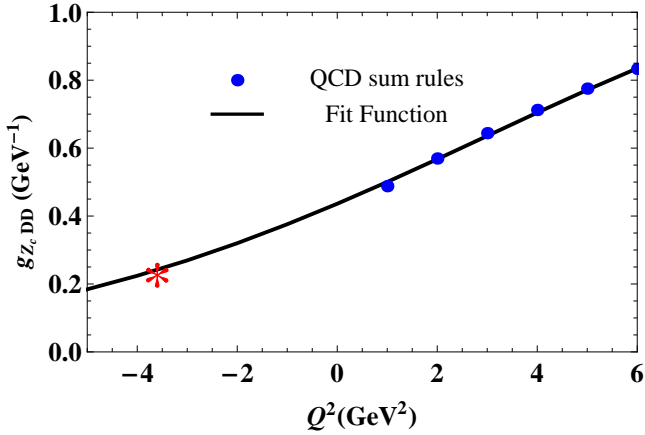}
\caption{The sum rule predictions and fit function for the strong coupling $%
g_{Z_{c}DD}(Q^{2})$. The star marks the point $Q^{2}=-m_{D^{0}}^{2}$. }
\label{fig:Fit}
\end{figure}

The width of the decay $Z_{c}\rightarrow D^{0}D^{-}$ depend on the strong
coupling at $D^{0}$ meson's mass shell. In other words, we need $%
g_{Z_{c}DD}(-m_{D^{0}}^{2})$ which cannot be accessed by direct sum rule
computations. Therefore,we use the fit function $\widetilde{F}(Q^{2})$ that
for the momenta $Q^{2}>0$ coincides with the sum rule results, and can be
easily extrapolated to the region of $Q^{2}<0$. The function (\ref{eq:FitF})
with the parameters $\widetilde{F}_{0}=0.44\ \mathrm{GeV}^{-1}$, $\widetilde{%
c}_{1}=2.38$ and $\widetilde{c}_{2}=-1.61$ meets these requirements. In
Fig.\ \ref{fig:Fit} we plot $\widetilde{F}(Q^{2})$ and the sum rule results
for $g_{Z_{c}DD}(Q^{2})$ demonstrating a very nice agreement between them.

At the mass shell $Q^{2}=-m_{D^{0}}^{2}$ the strong coupling is
\begin{equation}
g_{Z_{c}DD}(-m_{D^{0}}^{2})=(0.25\pm 0.05)\ \mathrm{GeV}^{-1}.
\label{eq:Coupl1}
\end{equation}%
The width of the decay $Z_{c}\rightarrow D^{0}D^{-}$ is calculated employing
the expression
\begin{equation}
\Gamma \lbrack Z_{c}\rightarrow D^{0}D^{-}]=\frac{g_{Z_{c}DD}^{2}m_{D}^{2}}{%
8\pi }\lambda \left( 1+\frac{\lambda ^{2}}{m_{D}^{2}}\right) ,  \label{eq:DW}
\end{equation}%
where $\lambda =\lambda \left( m,m_{D^{0}},m_{D}\right) .$

The partial width of this decay reads:%
\begin{equation}
\Gamma \lbrack Z_{c}\rightarrow D^{0}D^{-}]=(19\pm 5)~\mathrm{MeV}.
\label{eq:Width1}
\end{equation}%
It will be used below to estimate the total width of the tetraquark $Z_{c}$.

\section{Decay $Z_{c}\rightarrow J/\protect\psi \protect\rho ^{-}$}

\label{sec:Decay2}
The scalar tetraquark $Z_{c}$ in $S$-wave can also decay to the final state $%
J/\psi \rho ^{-}$. In the QCD light-cone sum rule approach this decay can be
explored through the correlation function%
\begin{equation}
\Pi _{\mu }(p,q)=i\int d^{4}xe^{ipx}\langle \rho (q)|\mathcal{T}\{J_{\mu
}^{J/\psi }(x)J^{\dagger }(0)\}|0\rangle ,  \label{eq:CorrF2}
\end{equation}%
where%
\begin{equation}
J_{\mu }^{J/\psi }(x)=\overline{c}_{i}(x)i\gamma _{\mu }c_{i}(x),
\label{eq:Curr2}
\end{equation}%
is the interpolating current for the vector meson $J/\psi $.

The correlation function $\Pi _{\mu }^{\mathrm{Phys}}(p,q)$ in terms of the
physical parameters of the tetraquark $Z_{c}$, and of the mesons $J/\psi $
and $\rho $ has the following form
\begin{eqnarray}
&&\Pi _{\mu }^{\mathrm{Phys}}(p,q)=\frac{\langle 0|J_{\mu }^{J/\psi }|J/\psi
\left( p\right) \rangle }{p^{2}-m_{J/\psi }^{2}}\langle J/\psi \left(
p\right) \rho (q)|Z_{c}(p^{\prime })\rangle  \notag \\
&&\times \frac{\langle Z_{c}(p^{\prime })|J^{\dagger }|0\rangle }{p^{\prime
2}-m^{2}}+\ldots ,  \label{eq:Phys3}
\end{eqnarray}%
where $m_{J/\psi }$ is the mass of the meson $J/\psi .$ In Eq.\ (\ref%
{eq:Phys3}) by the dots we denote contribution of the higher resonances and
continuum states. Here $p^{\prime }=p+q$ is the momentum of the tetraquark $%
Z_{c}$, where $p$ and $q$ are the momenta of the $J/\psi $ and $\rho $
mesons, respectively.

Further simplification of $\Pi _{\mu }^{\mathrm{Phys}}(p,q)$ can be achieved
by utilizing explicit expressions of the matrix elements $\langle 0|J_{\mu
}^{J/\psi }|J/\psi \left( p\right) \rangle ,$ $\langle Z_{c}(p^{\prime
})|J^{\dagger }|0\rangle $, and of the vertex $Z_{c}(p^{\prime })J/\psi
\left( p\right) \rho (q)$. The matrix element of the tetraquark $Z_{c}$ is
given by Eq.\ (\ref{eq:ME1}), whereas for the meson $J/\psi \left( p\right) $
we can use
\begin{equation}
\langle 0|J_{\mu }^{J/\psi }|J/\psi \left( p\right) \rangle =m_{J/\psi
}f_{J/\psi }\varepsilon _{\mu },  \label{eq:ME1A}
\end{equation}%
where $f_{J/\psi }$ and $\varepsilon _{\mu }$ are its decay constant and
polarization vector, respectively. We also model the three-state vertex as
\begin{eqnarray}
\langle J/\psi \left( p\right) \rho (q)|Z_{c}(p^{\prime })\rangle
&=&g_{Z_{c}J/\psi \rho }\left[ (p\cdot \varepsilon ^{\prime })(q\cdot
\varepsilon ^{\ast })\right.  \notag \\
&&-\left. (p\cdot q)(\varepsilon ^{\ast }\cdot \varepsilon ^{\prime })\right]
,  \label{eq:Vert}
\end{eqnarray}%
with $\varepsilon ^{\prime }$ being the polarization vector of the $\rho $%
-meson. Then $\Pi _{\mu }^{\mathrm{Phys}}(p,q)$ takes the form:%
\begin{eqnarray}
&&\Pi _{\mu }^{\mathrm{Phys}}(p,q)=g_{Z_{c}J/\psi \rho }\frac{m_{J/\psi
}f_{J/\psi }}{p^{2}-m_{J/\psi }^{2}}\frac{mf}{p^{\prime 2}-m^{2}}  \notag \\
&&\times \left[ \frac{1}{2}\left( m^{2}-m_{J/\psi }^{2}-q^{2}\right)
\varepsilon _{\mu }^{\prime }-p\cdot \varepsilon ^{\prime }q_{\mu }\right]
+\ldots .  \notag \\
&&  \label{eq:Phys2A}
\end{eqnarray}%
It contains different Lorentz structures $\sim \varepsilon _{\mu }^{\prime }$
and $q_{\mu }$. One of them should be chosen to fix the invariant amplitude
and carry out sum rule analysis. \ We choose the structure $\sim \varepsilon
_{\mu }^{\prime }$ and denote the corresponding invariant amplitude as $\Pi
^{\mathrm{Phys}}(p^{2},q^{2}).$

The second component of the sum rule is the correlation function $\Pi _{\mu
}(p,q)$ computed using quark propagators. For $\Pi _{\mu }^{\mathrm{OPE}%
}(p,q)$ we obtain
\begin{eqnarray}
&&\Pi _{\mu }^{\mathrm{OPE}}(p,q)=i^{2}\int d^{4}xe^{ipx}\epsilon \widetilde{%
\epsilon }\left[ \gamma _{5}\widetilde{S}_{c}^{aj}(x){}\gamma _{\mu }\right.
\notag \\
&&\left. \times \widetilde{S}_{c}^{ma}(-x){}\gamma _{5}\right] _{\alpha
\beta }\langle \rho (q)|\overline{d}_{\alpha }^{k}(0)u_{\beta
}^{n}(0)|0\rangle ,  \label{eq:CorrF3}
\end{eqnarray}%
where $\alpha $ and $\beta $ are the spinor indexes.

The expression for $\Pi _{\mu }^{\mathrm{OPE}}(p,q)$ can be written down in
a more detailed form. For these purposes, we \ first expand the local
operator $\overline{d}_{\alpha }^{a}u_{\beta }^{d}$ \ by means of the
formula
\begin{equation}
\overline{d}_{\alpha }^{a}u_{\beta }^{d}\rightarrow \frac{1}{4}\Gamma
_{\beta \alpha }^{j}\left( \overline{d}^{a}\Gamma ^{j}u^{d}\right) ,
\label{eq:MatEx}
\end{equation}%
with $\Gamma ^{j}$ being the full set of Dirac matrixes
\begin{equation*}
\Gamma ^{j}=\mathbf{1,\ }\gamma _{5},\ \gamma _{\lambda },\ i\gamma
_{5}\gamma _{\lambda },\ \sigma _{\lambda \rho }/\sqrt{2}.
\end{equation*}%
Applying the projector onto a color-singlet state $\delta ^{ad}/3$ we get
\begin{equation}
\overline{d}_{\alpha }^{a}u_{\beta }^{d}\rightarrow \frac{1}{12}\Gamma
_{\beta \alpha }^{j}\delta ^{ad}\left( \overline{d}\Gamma ^{j}u\right) ,
\label{eq:MatEl1}
\end{equation}%
where $\overline{d}\Gamma ^{j}u$ are the color-singlet local operators.
Substituting the last expression into Eq.\ (\ref{eq:CorrF3}) we see that the
correlation function $\Pi _{\mu }^{\mathrm{OPE}}(p,q)$ depends on the $\rho $%
-meson's two-particle local matrix elements. Some of them does not depend on
the $\rho $-meson momentum,

\begin{equation}
\langle 0|\overline{u}\gamma _{\mu }d|\rho (q,\lambda )\rangle =\epsilon
_{\mu }^{(\lambda )}f_{\rho }m_{\rho },  \label{eq:Melem1}
\end{equation}%
whereas others contain momentum factor

\begin{equation}
\langle 0|\overline{u}\sigma _{\mu \nu }d|\rho (q,\lambda )\rangle =if_{\rho
}^{T}(\epsilon _{\mu }^{(\lambda )}q_{\nu }-\epsilon _{\nu }^{(\lambda
)}q_{\mu }).  \label{eq:Melem1a}
\end{equation}%
There are also three-particle matrix elements that contribute to the
correlation function $\Pi _{\mu }^{\mathrm{OPE}}(p,q)$. They appear due to
insertion of gluon field strength tensor $G$ from the $c$-quark propagators
into the local operators $\overline{d}\Gamma ^{j}u.$ The $\rho $-meson
three-particle local matrix element
\begin{equation}
\langle 0|\overline{u}g\widetilde{G}_{\mu \nu }\gamma _{\nu }\gamma
_{5}d|\rho (q,\lambda )\rangle =f_{\rho }m_{\rho }^{3}\epsilon _{\mu
}^{(\lambda )}\zeta _{4\rho },  \label{eq:Melem2}
\end{equation}%
is a $q$ free quantity. But other matrix elements depend on the $\rho $%
-meson momentum

\begin{eqnarray}
\langle 0|\overline{u}gG_{\mu \nu }d|\rho (q,\lambda )\rangle &=&if_{\rho
}^{T}m_{\rho }^{3}\zeta _{4}^{T}(\epsilon _{\mu }^{(\lambda )}q_{\nu
}-\epsilon _{\nu }^{(\lambda )}q_{\mu }),  \notag \\
\langle 0|\overline{u}g\widetilde{G}_{\mu \nu }i\gamma _{5}d|\rho (q,\lambda
)\rangle &=&if_{\rho }^{T}m_{\rho }^{3}\widetilde{\zeta }_{4}^{T}(\epsilon
_{\mu }^{(\lambda )}q_{\nu }-\epsilon _{\nu }^{(\lambda )}q_{\mu }).  \notag
\\
&&  \label{eq:Melem2a}
\end{eqnarray}%
As a result, the correlation function contains only local matrix elements of
the $\rho $-meson and depends on the momenta $p$ and $q$. This is general
feature of QCD sum rules on the light-cone with a tetraquark and two
conventional mesons. Indeed, because a tetraquark contains four quarks,
after contracting two quark fields from its interpolating current with
relevant quarks from the interpolating current of a meson one gets a local
operator sandwiched between the vacuum and a second meson. The variety of
such local operators gives rise to different local matrix elements of the
meson rather that to its distribution amplitudes. Then the four-momentum
conservation in the tetraquark-meson-meson vertex requires setting $q=0$ (
for details, see Ref.\ \cite{Agaev:2016dev}). In the standard LCSR method
the choice $q=0$ is known as the soft-meson approximation \cite%
{Belyaev:1994zk}. At vertices composed of conventional mesons in general $%
q\neq 0$, and only in the soft-meson approximation one equates $q$ to zero,
whereas the tetraquark-meson-meson vertex can be analyzed in the context of
the LCSR method only if $q=0$. An important observation made in Ref.\ \cite%
{Belyaev:1994zk} is that the soft-meson approximation and full LCSR
treatment of the conventional mesons' vertices lead to results which
numerically are very close to each other. It is worth to note that the full
version of the sum rules on the light-cone is applicable to
tetraquark-tetraquark-meson vertices \cite{Agaev:2016srl}.

After substituting all aforementioned matrix elements into the expression of
the correlation function and performing the summation over color indices we
fix the local matrix elements of the $\rho $ meson that survive the soft
limit. It turns out that in the $q\rightarrow 0$ limit only the matrix
elements \ (\ref{eq:Melem1}) and \ (\ref{eq:Melem2}) contribute to the
invariant amplitude $\Pi ^{\mathrm{OPE}}(p^{2})$ [i.e. to $\Pi ^{\mathrm{OPE}%
}(p^{2},\ 0)$]. These matrix elements depend on the mass and decay constant
of the $\rho $-meson $m_{\rho }$, $f_{\rho }$, and on $\zeta _{4\rho }$
which normalizes the twist-4 matrix element of the $\rho $-meson \cite%
{Ball:1998ff}. The parameter $\zeta _{4\rho }$ was evaluated in the context
of QCD sum rule approach at the renormalization scale $\mu =1\,\,{\mathrm{GeV%
}}$ in Ref.\ \cite{Ball:2007zt} and is equal to $\zeta _{4\rho }=0.07\pm
0.03 $.

The Borel transform of the invariant amplitude $\Pi ^{\mathrm{OPE}}(p^{2})$
is given by the expression%
\begin{equation}
\Pi ^{\mathrm{OPE}}(M^{2})=\int_{4m_{c}^{2}}^{\infty }ds\widetilde{\rho }^{%
\mathrm{OPE}}(s)e^{-s/M^{2}},  \label{eq:OPE2}
\end{equation}%
where $\widetilde{\rho }^{\mathrm{OPE}}(s)$ is the corresponding spectral
density. In the present work we calculate $\widetilde{\rho }^{\mathrm{OPE}%
}(s)$ $\ $ by taking into account contribution of the condensates up to
dimension six. The spectral density has both the perturbative and
nonperturbative components
\begin{equation}
\widetilde{\rho }^{\mathrm{OPE}}(s)=\rho ^{\mathrm{pert.}}(s)+\rho ^{\mathrm{%
n.-pert.}}(s).  \label{eq:SD1}
\end{equation}%
After some computations for $\rho ^{\mathrm{pert.}}(s)$ we get
\begin{equation}
\rho ^{\mathrm{pert.}}(s)=\frac{f_{\rho }m_{\rho }(s+2m_{c}^{2})\sqrt{%
s(s-4m_{c}^{2})}}{24\pi ^{2}s}.  \label{eq:SD2}
\end{equation}%
The nonperturbative part of the spectral density $\rho ^{\mathrm{n.-pert.}%
}(s)$ contains terms proportional to the gluon condensates $\langle \alpha
_{s}G^{2}/\pi \rangle $, $\langle \alpha _{s}G^{2}/\pi \rangle ^{2}$ and $%
\langle g_{s}^{3}G^{3}\rangle $: Here we do not provide their explicit
expressions. The twist-4 contribution to $\Pi ^{\mathrm{OPE}}(M^{2})$ reads%
\begin{equation}
\Pi ^{\mathrm{OPE(tw4)}}(M^{2})=\frac{f_{\rho }m_{\rho }^{3}\zeta _{4\rho
}m_{c}^{2}}{8\pi }\int_{0}^{1}d\alpha \frac{e^{-m_{c}^{2}/M^{2}a(1-a)}}{%
a(1-a)}.  \label{eq:TW4}
\end{equation}

To derive the expression for the strong coupling $g_{Z_{c}J/\psi \rho }$ the
soft-meson approximation should be applied to the phenomenological side of
the sum rule as well. Because in the soft limit $p^{2}=p^{\prime 2}$, we
have to perform the Borel transformation of $\ \Pi ^{\mathrm{Phys}}(p^{2},0)$
over the variable $p^{2}$ and carry out calculations with one parameter $%
M^{2}$. To this end, we first transform $\Pi ^{\mathrm{Phys}}(p^{2},0)$ in
accordance with the prescription
\begin{equation}
\frac{1}{(p^{2}-m_{J/\psi }^{2})\left( p^{\prime 2}-m^{2}\right) }%
\rightarrow \frac{1}{(p^{2}-\widetilde{m}^{2})^{2}},  \label{eq:Doublepole}
\end{equation}%
where $\widetilde{m}^{2}=\left( m^{2}+m_{J/\psi }^{2}\right) /2$, \ and
instead of two terms with different poles get the double pole term. By
equating the physical and QCD sides and performing required manipulations we
get%
\begin{eqnarray}
&&\left( g_{Z_{c}J/\psi \rho }m_{J/\psi }f_{J/\psi }mf\frac{m^{2}-m_{J/\psi
}^{2}}{2}+AM^{2}\right)  \notag \\
&&\times \frac{e^{-\widetilde{m}^{2}/M^{2}}}{M^{2}}+\ldots =\Pi ^{\mathrm{OPE%
}}(M^{2}).  \label{eq:Equality1}
\end{eqnarray}

The equality given by Eq.\ (\ref{eq:Equality1}) is the master expression
which can be used to extract sum rule for the coupling $g_{Z_{c}J/\psi \rho
} $. It contain the term corresponding to the decay of the ground-state
tetraquark $Z_{c}$ and conventional contributions of higher resonances and
continuum states suppressed due to the Borel transformation; the latter is
denoted in Eq.\ (\ref{eq:Equality1}) by the dots. But in the soft limit
there are also terms $\sim A$ in the physical side which remain unsuppressed
even after the Borel transformation. They describe transition from the
excited states of the tetraquark $Z_{c}$ to the mesons $J/\psi \rho ^{-}$.
Of course, to obtain the final formula all contributions appearing as the
contamination should be removed from the physical side of the sum rule. The
situation with the ordinary suppressed terms is clear: they can be
subtracted from the correlation function $\Pi ^{\mathrm{OPE}}(M^{2})$ using
assumption on the quark-hadron duality. As a result the correlation function
acquires a dependence on the continuum threshold parameter $s_{0}$, i.e.,
becomes equal to $\Pi ^{\mathrm{OPE}}(M^{2},s_{0})$. The treatment of the
terms $\sim A$ requires some additional manipulations; they can be removed
by applying the operator \cite{Ioffe:1983ju}
\begin{equation}
\mathcal{P}(M^{2},\widetilde{m}^{2})=\left( 1-M^{2}\frac{d}{dM^{2}}\right)
M^{2}e^{\widetilde{m}^{2}/M^{2}},  \label{eq:Operator}
\end{equation}%
to both sides of Eq.\ (\ref{eq:Equality1}). Then the sum rule for the strong
coupling reads:%
\begin{eqnarray}
g_{Z_{c}J/\psi \rho } &=&\frac{2}{m_{J/\psi }f_{J/\psi }mf(m^{2}-m_{J/\psi
}^{2})}  \notag \\
&&\times \mathcal{P}(M^{2},\widetilde{m}^{2})\Pi ^{\mathrm{OPE}%
}(M^{2},s_{0}).  \label{eq:SR}
\end{eqnarray}%
The width of the decay $Z_{c}\rightarrow J/\psi \rho ^{-}$ can be calculated
using the formula%
\begin{eqnarray}
&&\Gamma \left( Z_{c}\rightarrow J/\psi \rho ^{-}\right) =\frac{%
g_{Z_{c}J/\psi \rho }^{2}m_{\rho }^{2}}{8\pi }\lambda \left( m,\ m_{J/\psi
},m_{\rho }\right)  \notag \\
&&\times \left[ 3+\frac{2\lambda ^{2}\left( m,\ m_{J/\psi },m_{\rho }\right)
}{m_{\rho }^{2}}\right] .  \label{eq:DW2}
\end{eqnarray}

In the sum rule (\ref{eq:SR}) for $M^{2}$ and $s_{0}$ we use the working
regions given by Eq.\ (\ref{eq:Wind}). For the strong coupling $%
g_{Z_{c}J/\psi \rho }$ we get%
\begin{equation}
g_{Z_{c}J/\psi \rho }=(0.56\pm 0.07)\ \mathrm{GeV}^{-1}.  \label{eq:SC1}
\end{equation}%
Then the width of the decay $Z_{c}\rightarrow J/\psi \rho ^{-}$ is
\begin{equation}
\Gamma \left[ Z_{c}\rightarrow J/\psi \rho ^{-}\right] =(15\pm 3)\ \mathrm{%
MeV.}  \label{eq:DW2num}
\end{equation}%
\ \qquad\ In accordance with our investigation, the total width of the
resonance $Z_{c}$ saturated by the dominant decay modes $Z_{c}\rightarrow
\eta _{c}(1S)\pi ^{-}$, $\eta _{c}(2S)\pi ^{-}$, $D^{0}D^{-}$ $\ $and $%
Z_{c}\rightarrow $ $J/\psi \rho ^{-}$ is
\begin{equation}
\Gamma =(147\pm 19)\ \mathrm{MeV.}  \label{eq:FW}
\end{equation}%
This is the second parameter of the resonance $Z_{c}$ to be compared with
the LHCb data; our result for the total with of $Z_{c}$ is in excellent
agreement with existing data $\Gamma =152_{-68}^{+83}\ \mathrm{MeV}$.


\section{Analysis and concluding remarks}

\label{sec:Conc}
We have performed quantitative analysis of the newly observed resonance $%
Z_{c}$ by calculating its spectroscopic parameters and total width. In
computations we have used different QCD sum rule approaches. Thus, the mass
and coupling of $Z_{c}$ have been evaluated by means of the two-point sum
rule method, whereas its decay channels have been analyzed using the
three-point and light-cone sum rule techniques.

We have calculated the spectroscopic parameters of the tetraquark $Z_{c}$
using the zero-width single-pole approximation. But the interpolating
current (\ref{eq:Curr}) couples also to the two-meson continuum $\eta
_{c}(1S)\pi ^{-}$, $\eta _{c}(2S)\pi ^{-}$, $J/\psi \rho ^{-}$, $D^{0}D^{-}$
and $D^{\ast 0}D^{\ast -}$ which can modify the results for $m$ and $f$
obtained in the present work. Effects of the two-meson continuum change the
zero-width approximation (\ref{eq:Modification}) and lead to the following
corrections \cite{Wang:2015nwa}%
\begin{equation}
\lambda ^{2}e^{-m^{2}/M^{2}}\rightarrow \lambda ^{2}\int_{\mathcal{M}%
^{2}}^{s_{0}}dsW(s)e^{-s/M^{2}}  \label{eq:Correction1}
\end{equation}%
and
\begin{equation}
\lambda ^{2}m^{2}e^{-m^{2}/M^{2}}\rightarrow \lambda ^{2}\int_{\mathcal{M}%
^{2}}^{s_{0}}dsW(s)se^{-s/M^{2}},  \label{eq:Correction2}
\end{equation}%
where $\lambda =mf$ and $\mathcal{M=}m_{D^{\ast 0}}+m_{D^{\ast -}}$. In
Eqs.\ (\ref{eq:Correction1}) and (\ref{eq:Correction2}) we have introduced
the weight function \
\begin{equation}
W(s)=\frac{1}{\pi }\frac{m\Gamma (s)}{\left( s-m^{2}\right) ^{2}+m^{2}\Gamma
^{2}(s)}  \label{eq:NewF1}
\end{equation}%
where
\begin{equation}
\Gamma (s)=\Gamma \frac{m}{s}\sqrt{\frac{s-\mathcal{M}^{2}}{m^{2}-\mathcal{M}%
^{2}}}.  \label{eq:NewF2}
\end{equation}%
Utilizing the central values of the $m$ and $\Gamma $, as well as $M^{2}=5\
\mathrm{GeV}^{2}$ and $s_{0}=20\ \mathrm{GeV}^{2}$, it is not difficult to
find that%
\begin{equation}
\lambda ^{2}\rightarrow 0.903\lambda ^{2}\rightarrow (0.95f)^{2}m^{2},
\label{eq:Rescale1}
\end{equation}%
and
\begin{equation}
\lambda ^{2}m^{2}\rightarrow 0.91\lambda ^{2}m^{2}\rightarrow
(0.955f)^{2}m^{4}.  \label{eq:Rescale2}
\end{equation}%
As is seen the two-meson effects result in rescaling $f\rightarrow 0.95f$ \
which changes it approximately by $5\%$ relative to its central value. These
effects are smaller than theoretical errors of the sum rule computations
themselves.

We have saturated the total width of the resonance $Z_{c}$ by its four
dominant decay modes $Z_{c}\rightarrow \eta _{c}(1S)\pi ^{-}$, $\eta
_{c}(2S)\pi ^{-}$, $D^{0}D^{-}$ and $J/\psi \rho ^{-}$. To calculate partial
widths of these decay channels we used two approaches in the framework of
the QCD sum rule method. Thus the decays $Z_{c}\rightarrow \eta _{c}(1S)\pi
^{-}$, $Z_{c}\rightarrow \eta _{c}(2S)\pi ^{-}$ and $Z_{c}\rightarrow
D^{0}D^{-}$ have been studied by applying three-point sum rules, whereas the
process $Z_{c}\rightarrow J/\psi \rho ^{-}$ has been investigated using the
LCSR method and soft-meson approximation. Predictions obtained for partial
widths of these $S$-wave decay channels have been used to evaluate the total
width of the resonance $Z_{c}$.

Our results for the mass $m=(4080~\pm 150)~\mathrm{MeV}$ and total width $%
\Gamma =(147\pm 19)\ \mathrm{MeV}$ of the resonance $Z_{c}$ are in a very
nice aggrement with experimental data of the LHCb Collaboration. This allows
us to interpret the new charged resonance as the scalar diquark-antidiquark
state with $cd\overline{c}\overline{u}$ content and $C\gamma _{5}\otimes
\gamma _{5}C$ structure. It presumably belongs to one of the charged $Z$%
-resonance multiplets, axial-vector members of which are the tetraquarks $%
Z_{c}^{\pm }(3900)$ and $Z_{c}^{\pm }(4330)$, respectively. The charged
resonances $Z_{c}^{\pm }(4330)$ and $Z_{c}^{\pm }(3900)$ were observed in
the $\psi ^{\prime }\pi ^{\pm }$ and $J/\psi \pi ^{\pm }$ invariant mass
distributions, i.e. they dominantly decay to these particles. The neutral
resonance $Z_{c}^{0}(3900)$ was discovered in the process $%
e^{+}e^{-}\rightarrow \pi ^{0}\pi ^{0}J/\psi $. Because $J/\psi $ and $\psi
^{\prime }$ are vector mesons, and $\psi ^{\prime }$ is the radial
excitation of $J/\psi $, it is natural to suggest that $Z_{c}(4330)$ is the
excited state of $Z_{c}(3900)$. This suggestion was originally made in Ref.\
\cite{Maiani:2014}, and confirmed later by sum rule calculations. Then the
resonance $Z_{c}^{-}(4100)$ fixed in the $\eta _{c}(1S)\pi ^{-}$ channel can
be interpreted as a scalar counterpart of these axial-vector tetraquarks. It
is also reasonable to assume that the neutral member of this family $%
Z_{c}^{0}(4100)$ will be seen in the processes $e^{+}e^{-}\rightarrow \pi
^{0}\pi ^{0}\eta _{c}(1S)$ with dominantly $\pi ^{0}\pi ^{0}$ mesons rather
than $D\overline{D}$ ones at the final state.

\appendix*

\section{ The quark propagators and two-point spectral density $\protect\rho %
^{\mathrm{OPE}}(s)$}

\renewcommand{\theequation}{\Alph{section}.\arabic{equation}} \label{sec:App}
The light and heavy quark propagators are necessary to find QCD side of the
different correlation functions. In the present work we use the light quark
propagator $S_{q}^{ab}(x)$ which is given by the following formula
\begin{eqnarray}
&&S_{q}^{ab}(x)=i\delta _{ab}\frac{\slashed x}{2\pi ^{2}x^{4}}-\delta _{ab}%
\frac{m_{q}}{4\pi ^{2}x^{2}}-\delta _{ab}\frac{\langle \overline{q}q\rangle
}{12}  \notag \\
&&+i\delta _{ab}\frac{\slashed xm_{q}\langle \overline{q}q\rangle }{48}%
-\delta _{ab}\frac{x^{2}}{192}\langle \overline{q}g_{s}\sigma Gq\rangle
+i\delta _{ab}\frac{x^{2}\slashed xm_{q}}{1152}\langle \overline{q}%
g_{s}\sigma Gq\rangle  \notag \\
&&-i\frac{g_{s}G_{ab}^{\alpha \beta }}{32\pi ^{2}x^{2}}\left[ \slashed x{%
\sigma _{\alpha \beta }+\sigma _{\alpha \beta }}\slashed x\right] -i\delta
_{ab}\frac{x^{2}\slashed xg_{s}^{2}\langle \overline{q}q\rangle ^{2}}{7776}
\notag \\
&&-\delta _{ab}\frac{x^{4}\langle \overline{q}q\rangle \langle
g_{s}q^{2}G^{2}\rangle }{27648}+\ldots .  \label{eq:Prop1}
\end{eqnarray}%
For the heavy $Q=c$ quark we utilize the propagator $S_{Q}^{ab}(x)$:
\begin{eqnarray}
&&S_{Q}^{ab}(x)=i\int \frac{d^{4}k}{(2\pi )^{4}}e^{-ikx}\Bigg \{\frac{\delta
_{ab}\left( {\slashed k}+m_{Q}\right) }{k^{2}-m_{Q}^{2}}  \notag \\
&&-\frac{g_{s}G_{ab}^{\alpha \beta }}{4}\frac{\sigma _{\alpha \beta }\left( {%
\slashed k}+m_{Q}\right) +\left( {\slashed k}+m_{Q}\right) \sigma _{\alpha
\beta }}{(k^{2}-m_{Q}^{2})^{2}}  \notag \\
&&+\frac{g_{s}^{2}G^{2}}{12}\delta _{ab}m_{Q}\frac{k^{2}+m_{Q}{\slashed k}}{%
(k^{2}-m_{Q}^{2})^{4}}+\frac{g_{s}^{3}G^{3}}{48}\delta _{ab}\frac{\left( {%
\slashed k}+m_{Q}\right) }{(k^{2}-m_{Q}^{2})^{6}}  \notag \\
&&\times \left[ {\slashed k}\left( k^{2}-3m_{Q}^{2}\right) +2m_{Q}\left(
2k^{2}-m_{Q}^{2}\right) \right] \left( {\slashed k}+m_{Q}\right) +\ldots %
\Bigg \}.  \notag \\
&&{}  \label{eq:Prop2}
\end{eqnarray}%
In the expressions above
\begin{eqnarray*}
&&G_{ab}^{\alpha \beta }=G_{A}^{\alpha \beta
}t_{ab}^{A},\,\,~~G^{2}=G_{\alpha \beta }^{A}G_{\alpha \beta }^{A}, \\
&&G^{3}=\,\,f^{ABC}G_{\mu \nu }^{A}G_{\nu \delta }^{B}G_{\delta \mu }^{C},
\end{eqnarray*}%
where $a,\,b=1,2,3$ are color indices and $A,B,C=1,\,2\,\ldots 8$. Here $%
t^{A}=\lambda ^{A}/2$ , where $\lambda ^{A}$ are the Gell-Mann matrices, and
the gluon field strength tensor is fixed at $x=0$, i.e. $G_{\alpha \beta
}^{A}\equiv G_{\alpha \beta }^{A}(0)$.

The spectral density $\rho ^{\mathrm{OPE}}(s)$ has the perturbative and
nonperturbative components%
\begin{equation}
\rho ^{\mathrm{OPE}}(s)=\rho ^{\mathrm{pert.}}(s)+\sum\limits_{N=3}^{10}\rho
^{\mathrm{DimN}}(s).  \label{eq:A1}
\end{equation}%
These components are determined by means of the formulas%
\begin{equation}
\rho ^{\mathrm{pert.(DimN)}}(s)=\int_{0}^{1}d\alpha \int_{0}^{1-\alpha
}d\beta \rho ^{\mathrm{pert.(DimN)}}(s,\alpha ,\beta ),  \label{eq:A2}
\end{equation}%
where $\rho ^{\mathrm{pert.}}(s,\alpha ,\beta )$ and $\rho ^{\mathrm{DimN}%
}(s,\alpha ,\beta )$ depend on $s$ and also on the Feynman parameters $%
\alpha $ and $\beta $. These functions are given by the following
expressions:
\begin{widetext}
\begin{eqnarray}
&&\rho ^{\mathrm{pert.}}(s,\alpha ,\beta )=\frac{\alpha \beta }{3\cdot
2^{9}\pi ^{6}(1-\alpha -\beta )D^{8}}\left[ s\alpha \beta (\alpha +\beta
-1)-m_{c}^{2}A\right] ^{2}\left[ 19s^{2}\alpha ^{2}\beta ^{2}(1-\alpha
-\beta )^{2}+m_{c}^{4}A^{2}-14m_{c}^{2}s\alpha \beta \right.   \notag \\
&&\left. \times \left( \beta ^{4}+r^{2}+\beta ^{3}(3\alpha -2)+\alpha \beta
(2-5\alpha +3\alpha ^{2})+\beta ^{2}(1-5\alpha +4\alpha ^{2})\right) \right]
\Theta \left[ L(s,\alpha ,\beta )\right] ,  \label{eq:A2a} \\
&&\rho ^{\mathrm{Dim3}}(s,\alpha ,\beta )=\frac{m_{c}\langle \bar{q}q\rangle
(\alpha +\beta )}{2^{4}\pi ^{4}D^{5}}\left\{ 2s^{2}\alpha ^{2}\beta
^{2}(1-\alpha -\beta )^{2}+m_{c}^{4}A^{2}-3m_{c}^{2}s\alpha \beta \left[
\beta ^{4}+r^{2}+\beta ^{3}(3\alpha -2)\right. \right.   \notag \\
&&\left. \left. +\alpha \beta (2-5\alpha +3\alpha ^{2})+\beta ^{2}(1-5\alpha
+4\alpha ^{2})\right] \right\} \Theta \left[ L(s,\alpha ,\beta )\right] ,
\label{eq:A3}
\end{eqnarray}%
\begin{eqnarray}
&&\rho ^{\mathrm{Dim4}}(s,\alpha ,\beta )=\frac{\langle \alpha _{s}G^{2}/\pi
\rangle \alpha \beta (\alpha +\beta )}{9\cdot 2^{9}\pi ^{4}(\alpha +\beta
-1)D^{6}}\left\{ 72s^{2}\alpha ^{2}\beta ^{2}(\alpha +\beta
-1)^{3}+m_{c}^{4}D^{2}\left[ 5\beta ^{3}+9\beta \alpha (3\alpha -2)+9\beta
^{2}(3\alpha -1)+\alpha ^{2}(5\alpha -9)\right] \right.   \notag \\
&&-8m_{c}^{2}s\alpha \beta \left[ 7\beta ^{5}+r^{2}(7\alpha -9)+\beta
^{4}(34\alpha -23)+\beta ^{3}(61\alpha ^{2}-88\alpha +25)+2\alpha \beta
(17\alpha ^{3}-44\alpha ^{2}+36\alpha -9)\right.   \notag \\
&&\left. \left. +\beta ^{2}(61\alpha ^{3}-124\alpha ^{2}+72\alpha -9)\right]
\right\} \Theta \left[ L(s,\alpha ,\beta )\right] ,  \label{eq:A4}
\end{eqnarray}%
\begin{eqnarray}
&&\rho ^{\mathrm{Dim5}}(s,\alpha ,\beta )=\frac{m_{c}\langle \overline{q}%
g_{s}\sigma Gq\rangle (\alpha +\beta )(1-\alpha -\beta )}{2^{6}\pi ^{4}D^{4}}%
\left[ 3s\alpha \beta (1-\alpha -\beta )+2m_{c}^{2}A\right] \Theta \left[
L(s,\alpha ,\beta )\right] ,  \label{eq:A4a} \\
&&\rho ^{\mathrm{Dim6}}(s,\alpha ,\beta )=\frac{\langle \bar{q}q\rangle
^{2}m_{c}^{2}}{12\pi ^{2}}\Theta \left[ M(s,\alpha )\right] -\frac{\alpha
\beta \langle \bar{q}q\rangle ^{2}g_{s}^{2}}{324\pi ^{4}D^{5}}(\alpha +\beta
-1)^{2}\left[ 4s\alpha \beta (1-\alpha -\beta )+m_{c}^{2}A\right] \Theta %
\left[ L(s,\alpha ,\beta )\right]   \notag \\
&&+\frac{\langle g_{s}^{3}G^{3}\rangle }{15\cdot 2^{14}\pi ^{6}(\beta
-1)^{2}(\alpha +\beta -1)D^{7}}\left[ m_{c}^{2}R_{1}(\alpha ,\beta
)+sR_{2}(\alpha ,\beta )\right] \Theta \left[ L(s,\alpha ,\beta )\right]
\label{eq:A5}
\end{eqnarray}%
\begin{eqnarray}
&&\rho ^{\mathrm{Dim7}}(s,\alpha ,\beta )=\frac{\langle \alpha _{s}G^{2}/\pi
\rangle \langle \bar{q}q\rangle m_{c}(\alpha +\beta )}{9\cdot 2^{5}\pi
^{2}D^{4}}\left\{ \alpha \left[ -3\beta ^{4}+6\beta ^{3}(1-\alpha )-4r\alpha
^{2}+\beta ^{2}(-3+9\alpha ^{2}-8\alpha ^{2})+\alpha \beta (-3+8\alpha
-5\alpha ^{2})\right] \right.   \notag \\
&&\left. +\beta \left[ 4\beta ^{4}+8\beta ^{2}r-3r^{2}-\beta ^{3}(4+5\alpha
)-3\alpha \beta (1-3\alpha +2\alpha ^{2})\right] \right\} \Theta \left[
L(s,\alpha ,\beta )\right] -\frac{\langle \alpha _{s}G^{2}/\pi \rangle
\langle \bar{q}q\rangle m_{c}}{9\cdot 2^{5}\pi ^{2}}\Theta \left[ M(s,\alpha
)\right] ,  \label{eq:A5a} \\
&&\rho ^{\mathrm{Dim8}}(s,\alpha ,\beta )=\frac{\langle \alpha _{s}G^{2}/\pi
\rangle ^{2}}{3\cdot 2^{10}\pi ^{2}D^{4}}\alpha ^{2}\beta ^{2}(1-\alpha
-\beta )\Theta \left[ L(s,\alpha ,\beta )\right] ,  \label{eq:A6}
\end{eqnarray}%
\begin{eqnarray}
&&\rho ^{\mathrm{Dim9}}(s,\alpha ,\beta )=\frac{\langle
g_{s}^{3}G^{3}\rangle \langle \bar{q}q\rangle m_{c}}{15\cdot 2^{6}\pi
^{4}D^{6}(\beta -1)s}\left\{ \alpha \left[ -4\beta ^{6}+6\beta ^{7}-4\beta
^{8}+\beta ^{9}-6\beta ^{4}\alpha ^{4}(\alpha -2)+3\alpha ^{5}(\alpha
-1)^{4}+3\beta \alpha ^{4}(\alpha -1)^{3}\right. \right.   \notag \\
&&\left. \times (1+4\alpha )+3\beta ^{4}\alpha ^{4}(\alpha -1)^{2}(4+5\alpha
)+3\beta ^{3}\alpha ^{4}(\alpha ^{2}+5\alpha -6)+\beta ^{4}(1-3\alpha ^{4})
\right] +\beta \left[ 3\beta ^{9}+4\beta ^{3}\alpha ^{5}(\alpha -1)\right.
\notag \\
&&+6\beta ^{2}\alpha ^{5}(\alpha -1)^{2}+4\beta \alpha ^{5}(\alpha
-1)^{3}+\alpha ^{5}(\alpha -1)^{4}-3\beta ^{8}(4+\alpha )+\beta
^{5}(3+12\alpha -9\alpha ^{2})-3\beta ^{7}(\alpha ^{2}-4\alpha -6)  \notag \\
&&\left. \left. +3\beta ^{6}(3\alpha ^{2}-6\alpha -4)+\beta ^{4}\alpha
(\alpha ^{4}-3\alpha -3)\right] \right\} \Theta \left[ L(s,\alpha ,\beta )%
\right] ,  \label{eq:A7} \\
&&\rho ^{\mathrm{Dim10}}(s,\alpha ,\beta )=0.  \label{eq:A7a}
\end{eqnarray}%
In equations above $ \Theta $ stands for the unit-step function and we have introduced the following notations%
\begin{eqnarray}
A &=&\beta ^{3}-2\beta \alpha (1-\alpha )-\alpha ^{2}(1-\alpha )+\beta
^{2}(2\alpha -1),\ \ D=\beta ^{2}-(1-\alpha )(\alpha +\beta ),\ \ r=\alpha
(1-\alpha ),  \notag \\
L(s,\alpha ,\beta ) &=&(1-\beta )\left[ s\alpha \beta (1-\alpha -\beta
)+m_{c}^{2}A\right] /D^{2}\ ,\ M(s,\alpha )=s\alpha (1-\alpha )-m_{c}^{2}.
\label{eq:A8}
\end{eqnarray}%
For brevity, we have also used
\begin{eqnarray*}
&&R_{1}(\alpha ,\beta )=-16\beta ^{12}+\beta ^{11}(80-27\alpha )+\beta
^{10}(-160+135\alpha -23\alpha ^{2})+2\alpha ^{7}(\alpha -1)^{3}(2\alpha
^{2}-4\alpha -5)+\beta \alpha ^{6}(\alpha -1)^{3} \\
&&\times (24\alpha ^{2}-4\alpha -9)-2\beta ^{9}(\alpha ^{3}-49\alpha
^{2}+135\alpha -80)+\beta ^{2}\alpha ^{5}(\alpha -1)^{2}(50\alpha
^{3}-60\alpha ^{2}+13\alpha -12)+\beta ^{3}\alpha ^{4}(\alpha -1)^{2}\left(
83\alpha ^{3}\right.  \\
&&\left. -85\alpha ^{2}+60\alpha -11\right) +\beta ^{8}(9\alpha ^{4}+3\alpha
^{3}-162\alpha ^{2}+270\alpha -80)+\beta ^{7}(20\alpha ^{5}-48\alpha
^{4}+8\alpha ^{3}+128\alpha ^{2}-135\alpha +16) \\
&&+\beta ^{6}\alpha (51\alpha ^{5}-106\alpha ^{4}+101\alpha ^{3}-22\alpha
^{2}-47\alpha +27)+\beta ^{5}\alpha ^{2}(88\alpha ^{5}-217\alpha
^{4}+210\alpha ^{3}-105\alpha ^{2}+18\alpha +6) \\
&&+\beta ^{4}\alpha ^{3}(99\alpha ^{5}-290\alpha ^{4}+33\alpha
^{3}-194\alpha ^{2}+54\alpha -5),
\end{eqnarray*}%
and
\begin{eqnarray*}
&&R_{2}(\alpha ,\beta )=\alpha \beta \left[ 2\beta ^{10}+\beta ^{9}(7\alpha
-160)+\beta ^{8}(320-3\alpha -50\alpha ^{2})-2\alpha ^{5}(\alpha
-1)^{3}(3\alpha ^{2}-6\alpha -13)-2\beta ^{2}\alpha ^{4}(\alpha
-1)^{2}\right.  \\
&&\times (14\alpha ^{2}-13\alpha +62)+\beta ^{3}\alpha ^{4}(186-400\alpha
-273\alpha ^{2}-59\alpha ^{3})-\beta ^{7}(25\alpha ^{3}-200\alpha
^{2}+58\alpha +320) \\
&&-\beta \alpha ^{4}(\alpha -1)^{2}(30\alpha ^{3}+4\alpha ^{2}-35\alpha
-31)+\beta ^{6}(75\alpha ^{3}-300\alpha ^{2}+122\alpha +160)-\beta
^{5}\left( 37\alpha ^{5}-31\alpha ^{4}+75\alpha ^{3}-200\alpha ^{2}\right.
\\
&&\left. \left. +93\alpha +32\right) -\beta ^{4}\alpha (92\alpha
^{5}-216\alpha ^{4}+124\alpha ^{3}-25\alpha ^{2}+50\alpha
-25)\right] .
\end{eqnarray*}%
Let us note that some of the functions $\rho (s,\alpha ,\beta )$ depend on $s
$ both explicitly and through the unit-step functions $\Theta \left[ L(s,\alpha
,\beta )\right] $ and/or $\Theta \left[ M(s,\alpha )\right] $, whereas in
others this dependence is generated only by the unit-step functions.

\end{widetext}

\end{document}